\documentclass[a4paper, 12pt, notitlepage, oneside]{scrartcl}
\makeatletter

\usepackage{lmodern}

\usepackage[T1]{fontenc}
\usepackage[utf8]{inputenc}
\usepackage[onehalfspacing]{setspace}
\usepackage[headsepline]{scrlayer-scrpage}
\clearscrheadfoot 
\ohead{\pagemark}
\usepackage{layout}
\usepackage[english]{babel}
\usepackage[left=2cm,right=2cm,top=2.5cm,bottom=2.5cm]{geometry}

\usepackage[boxed, linesnumbered,noend]{algorithm2e}
\usepackage{adjustbox}
\usepackage{eurosym}
\usepackage{lscape}
\usepackage{nicefrac}
\usepackage{array}
\usepackage{paralist}
\usepackage{authblk}
\usepackage{graphicx}
\usepackage{booktabs}
\usepackage{tabularx}
\usepackage{placeins}
\usepackage{amsmath}
\usepackage{mathtools}
\usepackage{bm}
\usepackage{amssymb}
\usepackage{color}
\usepackage[acronym]{glossaries}
\usepackage{pgfplots}
\usepackage[normal]{threeparttable}
\usepackage{ulem}
\usepackage{natbib}
\usepackage[titletoc,title]{appendix}
\usepackage[hyphens]{url}
\usepackage{ellipsis}
\usepackage{breqn} 
\usepackage{chngcntr}  
\usepackage{enumerate}
\usepackage{theorem}
\usepackage{etoolbox}
\usepackage{multirow}
\usepackage{colortbl}
\usepackage{subcaption}
\usepackage{tikz}
\usetikzlibrary{arrows.meta}
\usetikzlibrary{backgrounds}
\usepackage{pgfplotstable}
\usetikzlibrary{pgfplots.statistics}
\usepackage{tkz-kiviat/latex/tkz-kiviat}
\usepackage{numprint} 
\usetikzlibrary{arrows}
\makeglossaries

\DeclareOldFontCommand{\it}{\normalfont\itshape}{\mathit}

\begin{document}
\newacronym{abk:aseag}{ASEAG}{'Aachener Straßenbahn und Energieversorgungs-AG'}
\newacronym[firstplural=battery electric vehicles (BEVs)]{abk:bev}{BEV}{battery electric vehicle}
\newacronym[firstplural=battery-electric buses (BEBs)]{abk:beb}{BEB}{battery electric bus}
\newacronym{abk:bss}{BSS}{battery swap station}
\newacronym{abk:co2}{$\text{CO}_\text{2}$}{carbon dioxide}
%
\newacronym[firstplural=depot charging buses (DBEBs)]{abk:beb-on}{DBEB}{depot charging bus}
\newacronym[firstplural=depot charging facilities (DCFs)]{abk:srf}{DCF}{depot charging facility}
\newacronym{abk:dod}{DoD}{depth of discharge}
\newacronym{abk:ftp}{FTP}{fleet transformation problem}
\newacronym{abk:clp}{CLP}{charging location problem}
\newacronym{abk:clsap}{CLSAP}{charging location sequence assignment problem}
\newacronym{abk:sap}{SAP}{sequence assignment problem}
\newacronym{abk:vsp}{VSP}{vehicle scheduling problem}
\newacronym{abk:evrptw}{EVRP-TW}{electric vehicle routing problem with time windows}
\newacronym{abk:evrp}{EVRP}{electric vehicle routing problem}
\newacronym{abk:eu}{EU}{European Union}
\newacronym{abk:evsp}{EVSP}{electric vehicle scheduling problem}
\newacronym{abk:elrp}{ELRP}{electric location routing problem}
\newacronym{abk:frlp}{FRLP}{flow refueling location problem}
\newacronym{abk:frlp-lfc}{FRLP-LFC}{flow-refueling-location-problem with Load-Flow-Control}
\newacronym[firstplural=network charging buses (NBEBs)]{abk:beb-oc}{NBEB}{network charging bus}
\newacronym{abk:ghg}{GHG}{greenhouse gas}
\newacronym{abk:hpc}{HPC}{high-power-charging}
\newacronym[firstplural=internal combustion engine buses (ICEBs)]{abk:iceb}{ICEB}{internal combustion engine bus}
%
%
%
\newacronym{abk:mdvsp}{MDVSP}{multi-depot vehicle scheduling problem}
\newacronym{abk:mcfp}{MCFP}{minimum-cost flow problem}
\newacronym{abk:mip}{MIP}{mixed integer problem}
\newacronym{abk:nox}{$\text{NO}_\text{x}$}{nitrogen oxide}
\newacronym{abk:no2}{$\text{NO}_\text{2}$}{nitrogen dioxide}
\newacronym{abk:od}{OD-pair}{origin and destination pair}
\newacronym{abk:oem}{OEM}{original equipment manufacturer}
\newacronym{abk:pm}{$\text{PM}$}{particulate matter}
\newacronym{abk:pm2.5}{$\text{PM}_{2.5}$}{particulate matter ($2.5\,\mu m$)}
\newacronym{abk:pm10}{$\text{PM}_{10}$}{particulate matter ($10\,\mu m$)}
\newacronym{abk:pto}{PTO}{public transport operator}
%
\newacronym[firstplural=charging facilities (CFs)]{abk:rf}{CF}{charging facility}
\newacronym{abk:soc}{SOC}{state-of-charge}
\newacronym{abk:sdvsp}{SDVSP}{single-depot vehicle scheduling problem}
\newacronym{abk:tco}{TCO}{total cost of ownership}
\newacronym[firstplural=network charging facilities (NCFs)]{abk:frf}{NCF}{network charging facility}
\newacronym{abk:fs}{TS}{first or finale station of a bus line}
%
\newacronym{abk:vsplpr}{VSPLPR}{vehicle scheduling problem with length of path considerations}
%
%
%


\definecolor{rwth}   {RGB}{  0  84 159}
\definecolor{rwth-75}{RGB}{ 64 127 183}
\definecolor{rwth-50}{RGB}{142 186 229}
\definecolor{rwth-25}{RGB}{199 221 242}
\definecolor{rwth-10}{RGB}{232 241 250}

\definecolor{black}   {RGB}{  0   0   0}
\definecolor{black-75}{RGB}{100 101 103}
\definecolor{black-50}{RGB}{156 158 159}
\definecolor{black-25}{RGB}{207 209 210}
\definecolor{black-10}{RGB}{236 237 237}

\definecolor{magenta}   {RGB}{227   0 102}
\definecolor{magenta-75}{RGB}{233  96 136}
\definecolor{magenta-50}{RGB}{241 158 177}
\definecolor{magenta-25}{RGB}{249 210 218}
\definecolor{magenta-10}{RGB}{253 238 240}

\definecolor{yellow}   {RGB}{255 237   0}
\definecolor{yellow-75}{RGB}{255 240  85}
\definecolor{yellow-50}{RGB}{255 245 155}
\definecolor{yellow-25}{RGB}{255 250 209}
\definecolor{yellow-10}{RGB}{255 253 238}

\definecolor{petrol}   {RGB}{  0  97 101}
\definecolor{petrol-75}{RGB}{ 45 127 131}
\definecolor{petrol-50}{RGB}{125 164 167}
\definecolor{petrol-25}{RGB}{191 208 209}
\definecolor{petrol-10}{RGB}{230 236 236}

\definecolor{turkis}   {RGB}{  0 152 161}
\definecolor{turkis-75}{RGB}{  0 177 183}
\definecolor{turkis-50}{RGB}{137 204 207}
\definecolor{turkis-25}{RGB}{202 231 231}
\definecolor{turkis-10}{RGB}{235 246 246}

\definecolor{grun}   {RGB}{ 87 171  39}
\definecolor{grun-75}{RGB}{141 192  96}
\definecolor{grun-50}{RGB}{184 214 152}
\definecolor{grun-25}{RGB}{221 235 206}
\definecolor{grun-10}{RGB}{242 247 236}

\definecolor{maigrun}   {RGB}{189 205   0}
\definecolor{maigrun-75}{RGB}{208 217  92}
\definecolor{maigrun-50}{RGB}{224 230 154}
\definecolor{maigrun-25}{RGB}{240 243 208}
\definecolor{maigrun-10}{RGB}{249 250 237}

\definecolor{orange}   {RGB}{246 168   0}
\definecolor{orange-75}{RGB}{250 190  80}
\definecolor{orange-50}{RGB}{253 212 143}
\definecolor{orange-25}{RGB}{254 234 201}
\definecolor{orange-10}{RGB}{255 247 234}

\definecolor{rot}   {RGB}{204   7  30}
\definecolor{rot-75}{RGB}{216  92  65}
\definecolor{rot-50}{RGB}{230 150 121}
\definecolor{rot-25}{RGB}{243 205 187}
\definecolor{rot-10}{RGB}{250 235 227}

\definecolor{bordeaux}   {RGB}{161  16  53}
\definecolor{bordeaux-75}{RGB}{182  82  86}
\definecolor{bordeaux-50}{RGB}{205 139 135}
\definecolor{bordeaux-25}{RGB}{229 197 192}
\definecolor{bordeaux-10}{RGB}{245 232 229}

\definecolor{violett}   {RGB}{ 97  33  88}
\definecolor{violett-75}{RGB}{131  78 117}
\definecolor{violett-50}{RGB}{168 133 158}
\definecolor{violett-25}{RGB}{210 192 205}
\definecolor{violett-10}{RGB}{237 229 234}

\definecolor{lila}   {RGB}{122 111 172}
\definecolor{lila-75}{RGB}{155 145 193}
\definecolor{lila-50}{RGB}{188 181 215}
\definecolor{lila-25}{RGB}{222 218 235}
\definecolor{lila-10}{RGB}{242 240 247}

\title{\large A Concise Guide on the Integration of Battery Electric Buses\\into Urban Bus Networks}

\author[1]{\normalsize Nicolas Dirks\footnote{corresponding author}}
\author[2]{Dennis Wagner}
\author[3]{\normalsize Maximilian Schiffer}
\author[1]{\normalsize Grit Walther}

	\small
	\affil[1]{Chair of Operations Management, School of Business and Economics, RWTH Aachen University, Aachen  D-52072, Germany}
	\affil[2]{Daimler AG, EvoBus GmbH, Stuttgart D-70327, Germany}
	\affil[3]{TUM School of Management, Technical University of Munich, Munich D-80333, Germany}

	\affil[ ]{
	\scriptsize
	nicolas.dirks@om.rwth-aachen.de,
	dennis.d.wagner@daimler.com,
	schiffer@tum.de,
	walther@om.rwth-aachen.de}

\date{}

\maketitle
\begin{abstract}
\begin{singlespace}
{\small\noindent With the increasing market penetration of battery-electric buses into urban bus networks, practitioners face many novel planning problems. As a result, the interest in optimization-based decision-making for these planning problems increases but practitioners' requirements on planning solutions and current academic approaches often diverge. Against this background, this survey aims to provide a concise guide on optimization-based planning approaches for integrating battery-electric buses into urban bus networks for both practitioners and academics. First, we derive practitioners' requirements for integrating battery-electric buses from state-of-the-art specifications, project reports, and expert knowledge. Second, we analyze whether existing optimization-based planning models fulfill these practitioners' requirements. Based on this analysis, we carve out the existing gap between practice and research and discuss how to address these in future research.

\smallskip}
\end{singlespace}
\end{abstract}
\newpage
\section{Introduction}\label{sect:introduction}
Cities all around the world struggle with poor air quality, partially caused by transportation. To this end, \glspl{abk:beb} constitute a promising solution to improve urban air quality due to their local zero emissions. As a result, the market penetration of \glspl{abk:beb} has steadily increased in the past years (see Figure~\ref{fig:buses}). As Table~\ref{tab:projects} shows, cities all around the world started to integrate \glspl{abk:beb} into their public transportation network and many other cities plan to do so (cf.~\citealt{ZeEUSProject2017a}, \citealt{ELIPICT2018a}, \citealt{Danish2020a}). Consequently, practitioners are increasingly concerned with novel planning tasks and decision-making related to the deployment of \glspl{abk:beb} into existing fleets. 

When integrating \glspl{abk:beb}, complex planning decisions arise for fleet operators, who must consider both strategic and operational aspects simultaneously. At strategic level, operators must decide on the bus fleet's transformation and the installation of sufficient charging infrastructure. Herein, interdependencies and compatibilities between on-board charging devices of \glspl{abk:beb} and charging power levels of charging stations further complicate planning decisions. At operational level, fleet operations planning must ensure the bus timetable's operational feasibility under minimal costs. Accordingly, integrating \glspl{abk:beb} into urban bus networks requires advanced optimization-based planning models that can cope with these complex and interdependent planning decisions.

Academics respond to these planning requirements from practice by developing optimization-based planning models, and the number of scientific publications on planning models for \glspl{abk:beb} has increased considerably (see Figure~\ref{fig:pub}). However, due to the high inherent complexity of planning tasks, state-of-the-art optimization-based planning models often rely on simplifying assumptions, e.g., regarding (partial) recharging procedures, vehicle scheduling, or energy consumption, although these aspects can be of great importance for practitioners. For example, a simplifying assumption on the energy consumption may not be cost-efficient as buses get equipped with oversized battery capacities or may even lead to infeasible schedules as buses may run out of energy during real-world operations.
\begin{figure}[!hb]
	\pgfplotstableread[col sep = semicolon]{text/pgf/figBuses.csv}\DataA 
	\begin{minipage}{0.52\textwidth}
		\centering
		\begin{tikzpicture}
		\begin{axis}[
		tick label style={font=\small},
		x tick label style={rotate=315, anchor=west, align=left, font=\small},
		width=70mm,
		height=60mm,
		xlabel near ticks,
		ylabel near ticks,
		ytick pos=left,
		ylabel near ticks,
		tick style = {black},
		axis on top=true,
		xtick pos=left,
		ybar,
		xticklabels={,,
			2013,
			2014,
			2015,
			2016,
			2017,
			2018,
			2019,
			2020,
		},
		xtick distance={1},
		ylabel={Number of BEBs},
		xlabel={Year},
		ymin=0,ymax=4000,
		]
		\addplot[fill=rwth,draw=black,ultra thin] table [x=year,y=num]  {\DataA};
		\end{axis}
		\begin{axis}[
		tick label style={font=\small},
		width=70mm,
		height=60mm,
		axis y line=right,
		ylabel={Market penetration [\%]},
		yticklabels={,0,0.5,1.0,1.5,2.0,2.5},
		enlarge y limits=false,
		xtick distance=1,
		hide x axis,
		legend style={at={(0,1)},anchor=north west},
		legend cell align={left},
		ymin=0,ymax=12,
		]
		\addlegendimage{fill=rwth,mark=square*,only marks, mark size = 3.0, draw = black, ultra thin}
		\addlegendentry{per year}
		\addlegendimage{mark=*,only marks,sharp plot}
		\addlegendentry{3 years moving avg.}
		\end{axis}
		\end{tikzpicture}
		\caption{Number and market penetration of \glspl{abk:beb} in the European Union from 2013 to 2020, assuming a market volume of 160,000 buses, based on \citealt{EAFO2020a}.}
		\label{fig:buses}
	\end{minipage}
	\hfill
	\begin{minipage}{0.45\textwidth}
		\centering
		\pgfplotstableread[col sep = semicolon]{text/pgf/figPubA.csv}\DataA 
		\pgfplotstableread[col sep = semicolon]{text/pgf/figPubB.csv}\DataB 
		\begin{tikzpicture}
		\begin{axis}[
		tick label style={font=\small},
		x tick label style={rotate=315, anchor=west, align=left, font=\small},	
		width=70mm,
		height=60mm,
		xlabel near ticks,
		ylabel near ticks,
		legend style={at={(0,1)},anchor=north west,
			/tikz/every odd column/.style={yshift=-7pt},	
		},
		legend cell align={left},
		ytick pos=left,
		ylabel near ticks,
		tick style = {black},
		axis on top=true,
		xtick pos=left,
		ybar,
		xtick distance=1,
		xticklabels={,,
			2013,
			2014,
			2015,
			2016,
			2017,
			2018,
			2019,
			2020,
		},
		ylabel={Number of publications},
		xlabel={Year},
		ymin=0,ymax=12,
		]
		\addplot[fill=rwth,draw=black,ultra thin] table [x=year,y=num]  {\DataA};
		\addplot[black,mark=*,only marks,sharp plot] table [x=year,y=avg] {\DataB};
		\end{axis}
		\begin{axis}[
		width=70mm,
		height=60mm,
		enlarge y limits=false,
		xtick distance=1,
		hide x axis,
		hide y axis,
		legend style={at={(0,1)},anchor=north west,font=\small},
		legend cell align={left},
		ymin=0,ymax=12,
		xmin=2013,xmax=2020,
		]
		\addlegendimage{fill=rwth,mark=square*,only marks, mark size = 3.0, draw = black, ultra thin}
		\addlegendentry{per year}
		\addlegendimage{mark=*,only marks,sharp plot}
		\addlegendentry{3 years moving avg.}
		\end{axis}
		\end{tikzpicture}
		\caption{Number and three years moving average of publications on optimization-based planning for \glspl{abk:beb} between 2013 and 2020.}
		\label{fig:pub}
	\end{minipage}
\end{figure}
\begin{table*}[!ht]
	\centering
	\begin{threeparttable}
		\centering
		\newcommand{\PreserveBackslash}[1]{\let\temp=\\#1\let\\=\temp}
		\newcolumntype{C}[1]{>{\PreserveBackslash\centering}p{#1}}
		\setlength{\abovecaptionskip}{0.5ex}
		\caption{Number of conventional buses and \glspl{abk:beb} in cities all over the world.}
		\label{tab:projects}
		\footnotesize
		\def\c{\checkmark}
		\singlespacing
		\begin{tabular}{llll}
			\toprule
			City* & \#\,\glspl{abk:beb} & \#\,Buses & Reference\\
			\midrule
			Antelope Valley, USA & 49 & 88 & \citealt{AVTA2020a}\\
			Amsterdam, NL & 31 & 203 & \citealt{GVB2020a}\\	
			Berlin, DE & 121 & 1,440 & \citealt{BVG2018a}, \citealt{BVG2020a}\\
			Cluj-Napoca, RO & 41 & 297 & \citealt{Varga2020a}\\		
			Hamburg, DE & 35 & 1,000 & \citealt{Hochbahn2020a}\\
			London, UK & 280 & 9,142 & \citealt{London2019a}, \citealt{London2020a}\\
			Madrid, ESP & 82 & 2,109 & \citealt{Madrid2020a}\\
			Manchester, UK & 32 & 750 & \citealt{Manchester2020a}, \citealt{Manchester2020b}\\
			Moscow, RU & 300 & $11,000^{\text{**}}$ & \citealt{Moscow2019a}, \citealt{RussellPublishing2020a}\\
			Nur-Sultan, KZ & 100 & 1,082 & \citealt{Nur-Sultan2018a}, \citealt{Nur-Sultan2020a}\\
			Santiago de Chile, CL & 285 & 2,430 & \citealt{S.d.Chile2019a}, \citealt{S.d.Chile2019b}\\
			Shenzhen, CN & 16,359 & 16,359 & \citealt{Li2019a}\\
			Vienna, AT & 7 & 450 & \citealt{Vienna2020a}, \citealt{Vienna2020b}\\
			\bottomrule		
		\end{tabular}				
		\begin{tablenotes}				
			\scriptsize
			\item[*] We refer to specific transport operators in cities (see references) so that overall data for the cities may differ. 
			\item[**] including trams
		\end{tablenotes}
		\centering
	\end{threeparttable}
\end{table*}

Against this background, we provide a concise guide on integrating \glspl{abk:beb} into urban bus networks for practitioners and researchers. Herein, we focus on all planning tasks that are particularly affected by the integration of \glspl{abk:beb}. Specifically, we focus on transforming a bus fleet, installing charging infrastructure, and ensuring a fleet's operationality. 
In contrast to the existing review of \citealt{Clairand2019a} that focuses on electric vehicles in public transportation from a power system perspective, we focus on carving out all modeling requirements which are relevant to practice from a transport optimization perspective.

The contribution of this paper is fourfold. First, we derive model requirements relevant for practitioners based on specifications of state-of-the-art \glspl{abk:beb}, project reports, and expert knowledge. Second, we identify various problem classes related to the integration of \glspl{abk:beb}: \glsentryfullpl{abk:ftp}, \glsentryfullpl{abk:clp}, \glsentryfullpl{abk:clsap}, \glsentryfullpl{abk:sap}, and \glsentryfullpl{abk:vsp}. We then analyze to which extend existing works within these problem classes fulfill the respective practitioners' requirements. 
Third, we analyze the scientific literature concerning methodological aspects and practitioners' requirements, present results, and discuss scientific developments over time. Based on these analyses, we finally identify research gaps and promising research perspectives. 

The remainder of this paper is as follows. In Section~\ref{sect:requirements}, we derive practitioners' requirements based on state-of-the-art specifications, project reports, and expert knowledge. In Section~\ref{sect:literature}, we analyze existing optimization-based planning models with regard to covered practitioners' requirements and applied solution methods to identify research gaps and perspectives in Section~\ref{sect:gaps}. Section~\ref{sect:conclusion} concludes this survey with a short summary of its main findings.

\section{Practitioners' Requirements}\label{sect:requirements}
In the following, we identify practitioners' requirements relevant for planning the integration of \glspl{abk:beb} into urban bus networks. Herein, we cluster the practitioners' requirements into the following categories:
total costs~(T),
bus fleet management~(F),
charging infrastructure~(I),
energy consumption~(E),
charging procedure~(C),
vehicle scheduling~(S), and
deadheading \& detours~(D).
We structure this section based on these categories. Moreover, Table~\ref{tab:requirements} gives an overview of all derived practitioners' requirements and sorts them into each specific category.

\paragraph{Total Costs (T)}
Although most countries subsidize their public transportation systems, \glspl{abk:pto} must ensure cost-efficient bus network design and operation. Costs associated with the design and operation of a bus network include not only the investments in (respectively the replacement of) buses, batteries, and charging stations, but also all subsequent expenditures associated with the respective technologies. For example, (electric) motors, batteries, onboard charging devices, and charging station devices have to be checked, repaired, or exchanged regularly. Consequently, when integrating \glspl{abk:beb}, practitioners aim at minimizing costs for buses~(T1), batteries~(T2), charging infrastructure~(T3), maintenance costs for buses and batteries~(T4), maintenance costs for charging infrastructure~(T5), as well as costs for fuel or energy~(T6), respectively (cf.~\citealt{Init2019a}, \citealt{ViriCity2019a}).

\paragraph{Fleet Management (F)}
As buses have a limited service life of about~12 years (cf.~\citealt{Aber2016a}, \citealt{Lajunen2018a}), \glspl{abk:pto} must replace buses regularly, such that it is necessary to consider fleet purchases and sales~(F1) within a multi-period planning horizon~(F2) when aiming at optimal planning for a fleet's transformation.

Bus fleets often consist of heterogeneous buses~(F3) that differ concerning drive train technologies and other specifications that allow for a variety of services, e.g., public transport and school buses (cf.~\citealt{Akiyama2001a}). Due to demand and weight restrictions, decisions on passenger capacities~(F4) of buses need to be accounted for separately, e.g., whether solo, articulated, or bi-articulated buses with different passenger capacities are required.

In addition to these aspects, 
\begin{table*}[!hb]
	\centering
	\begin{threeparttable}
		\centering
		\newcommand{\PreserveBackslash}[1]{\let\temp=\\#1\let\\=\temp}
		\newcolumntype{C}[1]{>{\PreserveBackslash\centering}p{#1}}
		\setlength{\abovecaptionskip}{0.5ex}
		\caption{Practitioners' requirements for planning the integration of \glspl{abk:beb}.}
		\label{tab:requirements}
		\footnotesize
		\def\c{\checkmark}
		\singlespacing
		\begin{tabular}{lllll}
			\toprule
			ID & Requirement && ID & Requirement\\
			\midrule	
			\textbf{T} & \textbf{Total Costs} && I3 & Heterogeneous charging power levels\\
			T1 & Buses && I4 & Charging stations per location\\
			T2 & Batteries && \textbf{E} & \textbf{Energy Consumption}\\
			T3 & Charging stations && E1 & Non-linear energy consumption\\
			T4 & Maintenance of buses and batteries && E2 & Heterogeneous energy consumption\\
			T5 & Maintenance of charging stations && E3 & Stochastic energy consumption\\
			T6 & Fuel or energy costs && \textbf{C} & \textbf{Charging Procedures}\\
			\textbf{F} & \textbf{Fleet Management} && C1 & Partial recharging\\
			F1 & Fleet purchases and sales && C2 & Non-linear recharging\\
			F2 & Multi-period planning horizon && \textbf{S} & \textbf{Scheduling}\\
			F3 & Heterogeneous bus types && S1 & Trip assignment\\
			F4 & Passenger capacity && S2 & Sequence assignment\\
			F5 & Heterogeneous \glspl{abk:beb} && \textbf{D} & \textbf{Deadheading and Detours}\\		
			\textbf{I} & \textbf{Infrastructure} && D1 & Deadheading \\
			I1 & Charging power level selection && D2 & Detours for recharging\\
			I2 & Locations of charging stations &&&\\
			\bottomrule		
		\end{tabular}
	\end{threeparttable}
\end{table*}
technical characteristics of \glspl{abk:beb} may also differ concerning battery materials and general specifications~(F5). Table~\ref{tab:buses} lists state-of-the-art \glspl{abk:beb} with specifications, while Table~\ref{tab:batteries} lists state-of-the-art batteries with specifications. Both tables highlight the heterogeneity of state-of-the-art  \glspl{abk:beb} and battery technologies. While \glspl{abk:beb} mainly differ in passenger and battery capacity, batteries differ in volume and mass density, as well as in their \mbox{c-rate}, i.e., the number of full charging processes per hour, and over their lifetime. Herein, not only a battery's technology, but several additional factors such as its utilization, the charging behavior, and its operating temperature affect the battery's lifetime (cf.~\citealt{Atal2018a}, \citealt{Wang2019a}, \citealt{Proterra2020a}, \citealt{Daimler2020a}). Regardless of the technology, the battery capacity degrades over time (cf.~\citealt{Pelletier2017a}). Technical solutions are in place to ensure operational robustness such that the usable battery energy remains constant, i.e., the \gls{abk:dod} is monitored and continuously adjusted (cf.~\citealt{Denomme2012a}, \citealt{Franca2018a}, \citealt{MAN2020a}).

\paragraph{Infrastructure (I)}
A suitable and sufficient charging infrastructure is necessary to allow for recharging \glspl{abk:beb}. Technically, charging stations can be located anywhere in the street network. Practitioners, however, usually limit potential locations and decide on a specific charging concept: (1) depot charging between operations or overnight, (2) terminal charging between trip operations at initial and end stations of bus lines, (3) charging at bus stops during trip operations, (4) ex-route charging at charging stations which are not located at bus network stops or depots, or (5) swapping batteries at depots, terminals, or ex-route stations.
Still, decisions have to be taken on a suitable charging power level~(I1) and charging locations~(I2); both directly affect operations regarding the quantity and distances of deadheading trips (i.e., trips without passengers), the available recharging times, and hence operational costs (cf.~\citealt{Optibus2019a}).
\begin{table*}[!hb]
	\centering
	\begin{threeparttable}
		\centering
		\newcommand{\PreserveBackslash}[1]{\let\temp=\\#1\let\\=\temp}
		\newcolumntype{C}[1]{>{\PreserveBackslash\centering}p{#1}}
		\setlength{\abovecaptionskip}{0.5ex}
		\caption{State-of-the-art \gls{abk:beb} specifications.}
		\label{tab:buses}
		\footnotesize
		\def\c{\checkmark}
		\singlespacing
		\begin{tabular}{rlll}
			\toprule
			Model & \# Pax & Battery &  Reference \\
			\midrule
			BYD,	eBus	& 65\,-\,150 &	260\,-\,380\,kWh	&  \citealt{Hug2015a}\\ 		
			Ebusco,	Elec.	Citybus	2.2	& 90\,-\,130 &	363\,-\,525\,kWh	& \citealt{Ebusco2020a}\\
			Irizar,	ie	bus	& 76\,-\,155 &	90\,-\,525\,kWh	&	\citealt{Irizar2020a}\\	
			MAN,	Lion's	City	E	&  90\,-\,120 &	480\,-\,640\,kWh	&	 \citealt{MAN2020a}\\
			Mercedes Benz,	eCitaro	& 80\,-\,145 &	292\,-\,441\,kWh	&	 \citealt{EvoBus2020a}\\	
			Scania,	Citywide	LF	Elec.	& 95 &	250\,kWh	& \citealt{Scania2020a}\\
			Solaris,	Urbunio	Elec.	& 69\,-\,121 &	60\,-\,550\,kWh	&	 \citealt{Tschakert2020a}\\	
			VDL,	Citea	SLF(A)	Elec.	& 92\,-\,145 &	85\,-\,420\,kWh	& \citealt{VDL2020a}\\
			Volvo,	7900	Elec.	& 95\,-\,150 &	150\,-\,396	\,kWh&	\citealt{Volvo2020a}\\ 	
			Yutong,	U	12	& 86\,-\,100 &	422\,kWh	& \citealt{Yutong2020a}\\	
			\bottomrule		
		\end{tabular}				
		\centering
	\end{threeparttable}
	\vspace{-0.3cm}
\end{table*}
\begin{table*}[!hb]
	\centering
	\begin{threeparttable}
		\centering
		\newcommand{\PreserveBackslash}[1]{\let\temp=\\#1\let\\=\temp}
		\newcolumntype{C}[1]{>{\PreserveBackslash\centering}p{#1}}
		\setlength{\abovecaptionskip}{0.5ex}
		\caption{State-of-the-art batteries and specifications.}
		\label{tab:batteries}
		\footnotesize
		\def\c{\checkmark}
		\singlespacing
		\begin{tabular}{lllllll}
			\toprule
			Technology & Mass density & Volume density & c-Rate\textsuperscript{**} & Lifetime & Reference\\ 
			\midrule
			NMC&
			105\,Wh/kg& 140\,Wh/$\text{dm}^{3}$ & 0.69 & 5\,a & \citealt{Akasol2020a}\\			
			LMP&
			140\,Wh/kg& 111\,Wh/$\text{dm}^{3}$ & 0.2 & 10\,a & \citealt{BlueSolutions2019a}\\
			LTO &
			53\,Wh/kg& 55\,Wh/$\text{dm}^{3}$ & 13.16 & 3-4\,a & \citealt{Impact2018a}\\
			\bottomrule
		\end{tabular}				
		\begin{tablenotes}				
			\scriptsize
			\item[*] NMC: lithium nickel manganese cobalt oxides, LMP: lithium polymer battery, LTO: lithium-titanate battery.
			\item[**] based on continuous power charge. 					
		\end{tablenotes}
		\centering
	\end{threeparttable}
\end{table*}

So far, suitable charger types for buses are plug-in chargers as well as top-down and bottom-up pantographs. Some projects also tested inductive charging (cf. \citealt{Chen2018a}, \citealt{Bi2018a}), but concerns about electromagnetic compatibility and efficiency when using high charging power levels remain unsolved. However, experts envision potential technological advances in this area in the future (cf. \citealt{Momentum2020}).
Apparently, the manufacturers differ relatively little in shown specifications, which indicates that a quasi-standard for charging power levels and types exists. As can be seen in Table~\ref{tab:stations}, the charging powers of plug-in CCS and pantographs differ notably. While all suppliers offer plug-in charging with CCS up to 150\,kW, which is the maximum power reachable without liquid-cooled cables (cf.~\citealt{PhoenixContact2017a}), phantographs provide a charging rate of up to 600\,kW.
Also, the charging power level of the charging stations needs to match with the batteries installed in the \glspl{abk:beb} (cf.~\citealt{ELIPICTb}).
Therefore, it is necessary to take heterogeneous charging power levels~(I3) into account in order to install a charging infrastructure that is compatible with a heterogeneous bus fleet.
For example, \glspl{abk:beb} that are supposed to charge at a bottom-up pantograph
require special on-board charging devices (cf.,~e.g.,~\citealt{ABB2020a}, \citealt{Siemens2020a}). In addition to higher specific costs (cf.~\citealt{ABB2020b}), these onboard charging devices may lead to tighter weight limitations for batteries and to increased energy consumption. However, they may save battery capacity, charging stations, or even buses (cf.~\citealt{Mathieu2018a}). Finally, the number of charging stations per location~(I4) is crucial to ensure that several buses may charge at the same time at the same location (cf.~\citealt{Optibus2019a}).

\paragraph{Energy Consumption (E)}
Besides the battery capacity, the energy consumption determines the driving range and recharging needs of \glspl{abk:beb}. In general, the energy consumption is not linear~(E1) to the covered distance but rather depends on characteristics such as the topography, the general climatic conditions, or the number of acceleration and deceleration processes due to bus stations or traffic lights (cf.~\citealt{Init2019a}). Moreover, the energy consumption is heterogeneous~(E2), as it differs between different types of buses and batteries. For example, there are notable weight differences between solo or articulated buses and high-temperature or ambient-temperature batteries (cf.~\citealt{CACTUS2015a}), as well as between different battery capacities, battery technologies, and charging devices. Besides, the energy consumption is stochastic~(E3), as it depends on uncertainties such as the driving behavior, traffic conditions, or daily temperatures (cf.~\citealt{Init2019a}).
\begin{table*}[!hb]
	\centering
	\begin{threeparttable}
		\centering
		\newcommand{\PreserveBackslash}[1]{\let\temp=\\#1\let\\=\temp}
		\newcolumntype{C}[1]{>{\PreserveBackslash\centering}p{#1}}
		\setlength{\abovecaptionskip}{0.5ex}
		\caption{State-of-the-art charger types and specifications.}
		\label{tab:stations}
		\footnotesize
		\def\c{\checkmark}
		\singlespacing
		\begin{tabular}{rlll}
			\toprule
			Type & Supplier & Charging\\
			\midrule
			Plug-In  & ABB & 24\,-\,150\,kW\\
			CCS & Ekoenergetyka & 20\,-\,150\,kW\\
			& Siemens & 30\,-\,150\,kW\\		
			Bottom-Up  & ABB &  150\,-\,600\,kW\\
			Pantograph & Ekoenergetyka & $\leq 400$\,kW\\
			& Siemens & 60\,-\,120\,kW\\		
			Top-Down  & ABB &  150\,-\,600\,kW\\
			Pantograph & Siemens & 150\,-\,600\,kW\\
			\bottomrule		
		\end{tabular}				
		\begin{tablenotes}				
			\scriptsize
			\item Data from: ABB: \citealt{ABB2020c}, Ekoenergetyka: \citealt{Ekoenergetyka2020a}, Siemens: \citealt{Siemens2020a}
		\end{tablenotes}
		\centering
	\end{threeparttable}
	\vspace{-0.3cm}
\end{table*}

\paragraph{Charging Procedures (C)}
Buses operate within tight vehicle schedules. Hence, when integrating \glspl{abk:beb}, practitioners face the challenge of ensuring that any \gls{abk:beb} operates all dedicated trips without running out of energy. To this end, the standard option is overnight depot charging that ensures a reliable operation but avoids civil works within urban areas (cf.~\citealt{Mathieu2018a}). The alternative option is to (partially) recharge energy during operation~(C1) at (fast) charging stations (cf.~\citealt{ELIPICTb}), e.g., to increase operational flexibility.
The benefits of such partial recharging when routing battery-electric vehicles under consideration of time windows have been discussed in \cite{Schiffer2017a}.
Also, non-linear recharging processes~(C2) must be considered because voltage drops due to electrochemical impedances increase with an increasing \gls{abk:soc} (cf.~\citealt{Koch2017a}).

\paragraph{Scheduling (S)}
In order to transport passengers, a \gls{abk:pto} must operate a considerable amount of trips. In a typical bus network, these trips are defined by a bus timetable, i.e., (ordered) bus stations and service times. Based on this timetable, practitioners aim at identifying a feasible schedule to operate all trips of the timetable under minimal costs. Here, trips or pre-grouped sequences of trips must be assigned to the fleet's buses while considering requirements such as battery and passenger capacities. Apparently, the importance of such a trip~(S1) or sequence~(S2) assignment, i.e., vehicle scheduling, even increases when using \glspl{abk:beb}, as the state of charge of a \gls{abk:beb} correlates to the trips operated by a specific \gls{abk:beb}.

\paragraph{Deadheading and Detours (D)}
Deadheading trips~(D1), i.e., trips without passengers, are part of the operation in most bus networks.
Traditionally, these deadheading trips are either used for depot pull-in and pull-out or to overcome spatial differences between the end station and the first station of two different subsequently operated bus lines.
In general, disadvantages of deadheading are higher operational costs, a lower convenience for drivers, and a rather complex vehicle scheduling. However, a significant advantage is the potentially lower number of buses required due to the more flexible vehicle scheduling (cf.~\citealt{Optibus2019a}). When \glspl{abk:beb} are integrated, the importance of deadheading increases, as additional detours for recharging~(D2) may be beneficial.\\

The variety of requirement categoriers shows that a multitude of planning requirements exists with strategic as well as operational implications. 
Moreover, many interdependencies have to be taken into account, resulting in complex planning tasks that practitioners have to solve.

\section{State-of-the-Art}\label{sect:literature}

This section reviews state-of-the-art optimization-based planning models that focus on integrating \glspl{abk:beb} into existing bus networks. We identified publications by screening the pertinent data banks, e.g., Web of Science, Google Scholar, ScienceDirect, with related keywords, e.g., electric buses, charging infrastructure, vehicle scheduling.
We then checked all publications cited by the publications found via our data bank search in a subsequent step. To keep this paper concise, we focus on publications that use an optimization-based modeling approach and are written in English.

We group the existing publications into five problem classes.
\Glspl{abk:ftp} focus on purchases and sales of buses with multi-period decisions. \Glspl{abk:clp} focus on finding suitable locations for charging stations. \Glspl{abk:sap} focus on an assignment of trip sequences to buses. \Glspl{abk:clsap} base decisions on charging station locations on a sequence assignment to consider operational feasibility. \Glspl{abk:vsp} optimize a trip assignment, i.e., a comprehensive vehicle scheduling.
Table~\ref{tab:classification} gives an overview of all problem classes and the respective nomenclature.

In the following, we provide a detailed analysis of all publications structured by problem classes as shown in Table~\ref{tab:classification} before we provide a comparison of problem classes in order to derive promising research perspectives in Section~\ref{sect:gaps}.

\subsection{Fleet Transformation Problems}\label{sub:ftp}

\glspl{abk:ftp} aim at identifying a cost-optimal and multi-period bus fleet transformation plan. In this context, decisions on the purchase of buses and the installation of charging stations are necessary. The models anticipate operational aspects and costs by using simplified assumptions. They either focus on depot charging (cf.~\citealt{Islam2019a}), on charging at terminals (cf.~\citealt{Li2018a}, \citealt{Dirks2021a}), or at bus stops (cf.~\citealt{Pelletier2019a}).

Table~\ref{tab:ftp} shows to which extend existing \glspl{abk:ftp} meet the practitioners' requirements.
By definition, all \glspl{abk:ftp} account for fleet purchases and sales~(F1) as well as for a multi-period planning horizon~(F2).
While in general \glspl{abk:ftp} account for many requirements in the field of total costs~(T) and bus fleet management~(F), they rarely cover requirements of the other categories.
The fact that requirements related to charging procedures~(C), as well as deadheading and detours~(D), are not covered at all indicates that state-of-the-art \glspl{abk:ftp} tend to neglect operational implications.
\begin{table*}[!hb]
	\centering
	\begin{threeparttable}
		\centering
		\newcommand{\PreserveBackslash}[1]{\let\temp=\\#1\let\\=\temp}
		\newcolumntype{C}[1]{>{\PreserveBackslash\centering}p{#1}}
		\setlength{\abovecaptionskip}{0.5ex}
		\caption{Problem classes.}
		\label{tab:classification}
		\footnotesize
		\def\c{\checkmark}
		\singlespacing
		\begin{tabular}{lll}
			\toprule
			Abbrev. & Description & Section\\
			\midrule	
			\acrshort{abk:ftp}	& \Acrlong{abk:ftp} & \ref{sub:ftp}\\
			\acrshort{abk:clp}	& \Acrlong{abk:clp} & \ref{sub:clp}\\
			\acrshort{abk:sap}   & \Acrlong{abk:sap} & \ref{sub:sap}\\
			\acrshort{abk:clsap} & \Acrlong{abk:clsap} & \ref{sub:clsap}\\
			\acrshort{abk:vsp}	& \Acrlong{abk:vsp} & \ref{sub:vsp}\\
			\bottomrule		
		\end{tabular}
	\end{threeparttable}
\end{table*}
\begin{table*}[!hb]
	\centering
	\begin{threeparttable}
		\centering
		\newcommand{\PreserveBackslash}[1]{\let\temp=\\#1\let\\=\temp}
		\newcolumntype{C}[1]{>{\PreserveBackslash\centering}p{#1}}
		\setlength{\abovecaptionskip}{0.5ex}
		\caption{Fulfillment of the practitioners' requirements by \glspl{abk:ftp}.}
		\label{tab:ftp}
		\footnotesize
		\def\c{\checkmark}
		\addtolength{\tabcolsep}{-3.5pt}
		\singlespacing
		\begin{tabular}{l|l|llllll|lllll|llll|lll|ll|ll|ll}
			\toprule
			&& \multicolumn{6}{c|}{\textbf{T}} & \multicolumn{5}{c|}{\textbf{F}} & \multicolumn{4}{c|}{\textbf{I}}
			& \multicolumn{3}{c|}{\textbf{E}} & \multicolumn{2}{c|}{\textbf{C}} & \multicolumn{2}{c|}{\textbf{S}} & \multicolumn{2}{c}{\textbf{D}}\\
			& charg. loc. & 1 & 2 & 3 & 4 & 5 & 6 &
			1 & 2 & 3 & 4 & 5 &
			1 & 2 & 3 & 4 &
			1 & 2 & 3 &
			1 & 2 & 1 & 2 & 1 & 2\\
			\midrule
			\citealt{Dirks2021a} &terminals&\c&\c&\c&\c&\c&\c&\c&\c&\c& &\c& &\c& & & & & &\c& & &\c&\c& \\
\citealt{Islam2019a} &depot&\c&\c&\c&\c&\c&\c&\c&\c&\c& & & & & & & & & & & & & & & \\
\citealt{Li2018a} &terminals&\c& & & & &\c&\c&\c&\c& & & & & & & & & & & & &\c& & \\
\citealt{Pelletier2019a} &stops&\c& &\c&\c& &\c&\c&\c&\c&\c&\c&\c& &\c& & &\c& & & & & & &

			\\\bottomrule
		\end{tabular}
		\begin{tablenotes}				
			\scriptsize
			\item T: total costs, F: bus fleet, I: charging infrastructure, E: energy consumption, C: charging procedures, S: vehicle scheduling, D: deadheading and detours.
		\end{tablenotes}
		\centering
	\end{threeparttable}
\end{table*}

\subsection{Charging Location Problems}\label{sub:clp}

\glspl{abk:clp} mainly aim at identifying charging station locations~(I2) either for single bus lines or for entire bus networks. Also, some publications focus on planning locations of large-scale charging parks.

Early publications often focused on models for solving charging location problems for single multi-stop bus lines
(cf.~\citealt{Chen2013a},
\citealt{Kunith2013a},
\citealt{Wehres2016a}, 
\citealt{Berthold2017a},
\citealt{Rohrbeck2018a}).
These planning models are often based on pilot projects which focused on electrifying single bus lines for testing purposes. All these publications allow partial recharging at in-line bus stops. By definition, vehicle scheduling, as well as deadheading and detours, are neglected. Models differ in whether they account for specific characteristics such as uncertainties in energy consumption (cf.~\citealt{Wehres2016a}), battery aging (cf.~\citealt{Rohrbeck2018a}), or a non-linear recharging behavior (cf.~\citealt{Kunith2013a}).

More recent \gls{abk:clp} publications introduced models that allow charging location planning for entire bus networks including multiple bus lines
(cf.~\citealt{Kunith2016a},
\citealt{Kunith2017b},
\citealt{Sebastiani2016a},
\citealt{Xylia2017a},
\citealt{Xylia2017b},
\citealt{Liu2018b},
\citealt{Lotfi2020a},
\citealt{Zhou2020b}).
When accounting for multiple bus lines, synergies between several bus lines can be leveraged to identify better solutions compared to considering bus lines separately, as buses that operate on different bus lines can use the same charging stations. However, these models still lack the synergies that arise from deadheading. They mainly differ in decisions on technical specifications, e.g., regarding the considered drivetrains, or in applied (technical) methodologies, e.g., regarding the calculation of energy consumption.

For megacities with a high public transport demand that operate large \gls{abk:beb} fleets, it may be reasonable to establish large charging facilities in addition to the depots, at which \glspl{abk:beb} may either recharge or swap batteries in between operations. We refer to those facilities as charging parks. Models focusing on charging parks have been developed and applied, especially in Asian or Pacific megacities (cf.~\citealt{Lin2019a}, \citealt{Lin2019b}, \citealt{An2019a}, \citealt{An2020a}). For simplicity, these models are based on aggregation methodologies, in which clusters of bus routes are assigned to charging parks, and thus the recharging demand is estimated.

Table~\ref{tab:clp} shows to which extend existing \glspl{abk:clp} fulfill the different practitioners' requirements.
By definition, all \glspl{abk:clp} decide on locations for charging stations~(I4) and their costs~(T3). However, many other requirements are neglected. While about one-third of the publications fulfill requirements in the fields of total costs~(T), charging infrastructure~(F), and charging processes~(C), only a few cover requirements related to bus fleet management~(F), energy consumption~(E), and deadheading and detours~(D). Also, by definition, no \gls{abk:clp} accounts for requirements related to vehicle scheduling~(S).

\subsection{Sequence Assignment Problems}\label{sub:sap}
\glspl{abk:sap} aim at finding an efficient vehicle schedule by predetermining feasible sequences of trips and assigning these to buses or bus types.
They either focus on depot charging (cf.~\citealt{Ke2016a}, \citealt{Rogge2018a}, \citealt{Yao2020a}), or assume a given (fast) charging infrastructure (cf. \citealt{Reuer2015a}, \citealt{Sassi2017a}).
All \gls{abk:sap} publications that focus on depot charging solve the assignment of trip sequences to buses with a genetic algorithm.
\cite{Sassi2017a} and \cite{Reuer2015a} allow for charging at terminals and use a greedy algorithm or a max-flow formulation based on \citet{Kliewer2006a} respectively. In addition, \citet{Rogge2018a} include a comprehensive energy consumption simulation based on \citet{Sinhuber2012a} with data on service trips, vehicle types, and route characteristics.

Table~\ref{tab:sap} shows to which extend existing \glspl{abk:sap} fulfill the different practitioners' requirements. By definition, all \glspl{abk:sap} assign trip sequences to buses~(S2).
In total, \glspl{abk:sap} cover many requirements relevant for practice, with requirements related to total costs~(T) and deadheading~(D) prevailing. Occasionally, some \glspl{abk:sap} comprise bus fleet management~(F), energy consumption~(E), and vehicle scheduling requirements. Contrarily, requirements in the field of charging infrastructure~(I) and charging procedure~(C) are less prevalent.
\begin{table*}[!hb]
	\centering
	\begin{threeparttable}
		\centering
		\newcommand{\PreserveBackslash}[1]{\let\temp=\\#1\let\\=\temp}
		\newcolumntype{C}[1]{>{\PreserveBackslash\centering}p{#1}}
		\setlength{\abovecaptionskip}{0.5ex}
		\caption{Fulfillment of the practitioners' requirements by \glspl{abk:clp}.}
		\label{tab:clp}
		\footnotesize
		\def\c{\checkmark}
		\addtolength{\tabcolsep}{-3.5pt}
		\singlespacing
		\begin{tabular}{l|l|llllll|lllll|llll|lll|ll|ll|ll}
			\toprule
			&& \multicolumn{6}{c|}{\textbf{T}} & \multicolumn{5}{c|}{\textbf{F}} & \multicolumn{4}{c|}{\textbf{I}}
			& \multicolumn{3}{c|}{\textbf{E}} & \multicolumn{2}{c|}{\textbf{C}} & \multicolumn{2}{c|}{\textbf{S}} & \multicolumn{2}{c}{\textbf{D}}\\
			& charg. loc. & 1 & 2 & 3 & 4 & 5 & 6 &
			1 & 2 & 3 & 4 & 5 &
			1 & 2 & 3 & 4 &
			1 & 2 & 3 &
			1 & 2 & 1 & 2 & 1 & 2\\
			\midrule
			\citealt{An2019a} &swapping& & &\c& & &\c& & &\c& & &\c&\c& &\c& & &\c& & & & &\c&\c\\
\citealt{An2020a} &stations&\c& &\c& &\c&\c& & &\c& & & &\c& & & & &\c&\c& & & &\c&\c\\
\citealt{Berthold2017a} &stops& & &\c& & & & & & & & & &\c& & & & & &\c& & & & & \\
\citealt{Chen2013a} &stops& &\c&\c& & &\c& & & & & &\c&\c&\c& & & & &\c& & & & & \\
\citealt{Kunith2013a} &stops& & &\c& & & & & & & & & &\c& & &\c& & &\c&\c& & & & \\
\citealt{Kunith2016a} &terminals& & &\c& & & & & & &\c& & &\c& &\c&\c& & &\c&\c& & & & \\
\citealt{Kunith2017b} &terminals& &\c&\c& & & & & & &\c&\c& &\c& &\c&\c& & &\c&\c& & & & \\
\citealt{Lin2019a} &stations& & &\c& &\c&\c& & & & & & &\c& &\c& & & & & & & &\c&\c\\
\citealt{Lin2019b} &depot& & &\c& &\c&\c& &\c& & & & &\c& &\c& & & & & & & & & \\
\citealt{Liu2018b} &stops& &\c&\c& & & & & &\c& &\c&\c&\c&\c& &\c&\c&\c&\c& & & & & \\
\citealt{Lotfi2020a} &terminals&\c&\c&\c&\c&\c&\c& & & & &\c&\c&\c&\c& & & & &\c& & & & & \\
\citealt{Rohrbeck2018a} &stops& &\c&\c& & & & &\c& & & & &\c& & & & & &\c& & & & & \\
\citealt{Sebastiani2016a} &stops& & &\c& & & & & & & & & &\c& & &\c& & &\c&\c& & & & \\
\citealt{Wehres2016a} &stops& &\c&\c& & & & & & & & &\c&\c&\c& & & &\c&\c& & & & & \\
\citealt{Xylia2017a} &stops& & &\c&\c&\c&\c& & &\c& & &\c&\c&\c& & & & &\c& & & & & \\
\citealt{Xylia2017b} &stops& & &\c&\c&\c&\c& & &\c& & &\c&\c&\c& & & & &\c& & & & & \\
\citealt{Zhou2020b} &terminals&\c& &\c& & & & & &\c& & & &\c& &\c& & & & & & & & &

			\\\bottomrule
		\end{tabular}				
		\begin{tablenotes}				
			\scriptsize
			\item T: total costs, F: bus fleet, I: charging infrastructure, E: energy consumption, C: charging procedures, S: vehicle scheduling, D: deadheading and detours.
		\end{tablenotes}
		\centering
	\end{threeparttable}
\end{table*}
\begin{table*}[!hb]
	\centering
	\begin{threeparttable}
		\centering
		\newcommand{\PreserveBackslash}[1]{\let\temp=\\#1\let\\=\temp}
		\newcolumntype{C}[1]{>{\PreserveBackslash\centering}p{#1}}
		\setlength{\abovecaptionskip}{0.5ex}
		\caption{Fulfillment of the practitioners' requirements by \glspl{abk:sap}.}
		\label{tab:sap}
		\footnotesize
		\def\c{\checkmark}
		\addtolength{\tabcolsep}{-3.5pt}
		\singlespacing
		\begin{tabular}{l|l|llllll|lllll|llll|lll|ll|ll|ll}
			\toprule
			&& \multicolumn{6}{c|}{\textbf{T}} & \multicolumn{5}{c|}{\textbf{F}} & \multicolumn{4}{c|}{\textbf{I}}
			& \multicolumn{3}{c|}{\textbf{E}} & \multicolumn{2}{c|}{\textbf{C}} & \multicolumn{2}{c|}{\textbf{S}} & \multicolumn{2}{c}{\textbf{D}}\\
			& charg. loc. & 1 & 2 & 3 & 4 & 5 & 6 &
			1 & 2 & 3 & 4 & 5 &
			1 & 2 & 3 & 4 &
			1 & 2 & 3 &
			1 & 2 & 1 & 2 & 1 & 2\\
			\midrule
			\citealt{Ke2016a} &depot&\c&\c&\c& & &\c& & & & &\c& & & &\c& &\c& &\c&\c& &\c&\c&\c\\
\citealt{Reuer2015a} &terminals&\c& & & & &\c& & &\c& & & & & & & & & & & & &\c&\c&\c\\
\citealt{Rogge2018a} &depot&\c&\c&\c&\c&\c&\c& & &\c&\c&\c& & & &\c&\c&\c& & & & &\c&\c&\c\\
\citealt{Sassi2017a} &terminals& & & & & &\c& & &\c& & & & & & & & & &\c& & &\c&\c& \\
\citealt{Yao2020a} &depot&\c&\c&\c& & &\c& & & &\c&\c& & & &\c& &\c& & & & &\c&\c&\c

			\\\bottomrule
		\end{tabular}				
		\begin{tablenotes}				
			\scriptsize
			\item T: total costs, F: bus fleet, I: charging infrastructure, E: energy consumption, C: charging procedures, S: vehicle scheduling, D: deadheading and detours.
		\end{tablenotes}
		\centering
	\end{threeparttable}
\end{table*}

\subsection{Charging Location Sequence Assignment Problems}\label{sub:clsap}
\glspl{abk:clsap} link decisions on charging station locations with operational sequence assignment, thus enabling these models to consider deadheading and to exploit synergies between multiple bus lines. However, charging locations are limited to depots as well as terminals and in-line bus stops.
\citet{Li2018c} \& \citet{Li2019a} use a time-space-energy network, predetermining all potential trip sequences a-priori. Here, charging stations are installed in dependence of selected sequences.
Additionally, the authors consider passenger flows within a time-space network. \citet{Wei2018a} introduce a model for both replacing \glspl{abk:iceb} by \glspl{abk:beb} and installing charging stations at terminal bus stops. Their model considers deadheading but no partial recharging.
\citet{Jefferies2020a} combine optimization and discrete-event simulation in order to plan a cost-efficient and feasible bus network with \glspl{abk:beb} that charge on opportunity. While the simulation considers complex relations such as temperature-dependent energy consumption, the optimization bases on worst-case parameters.

Table~\ref{tab:clsap} shows to which extend existing \glspl{abk:clsap} fulfill the different practitioners' requirements.
By definition, all \glspl{abk:clsap} not only decide on locations for charging stations~(I4) and their costs~(T3), but they also assign trip sequences to buses~(S2).

\subsection{Vehicle Scheduling Problems}\label{sub:vsp}
The main goal of \glspl{abk:vsp} is to compute an optimal vehicle schedule that requires a minimum number of buses with limited driving ranges for a given charging infrastructure
(cf.~\citealt{Li2014a},
\citealt{Paul2014a},
\citealt{Wen2016a},
\citealt{Adler2017a},
\citealt{vanKootenNiekerk2017a},
\citealt{Janovec2019a},
\citealt{Rinaldi2019a},
\citealt{Rinaldi2019b},
\citealt{Tang2019a},
\citealt{Liu2020a},
\citealt{Teng2020a}).
For example, models differ with respect to the considered charging staton concept, e.g., whether charging stations are exclusively located at terminal or at in-line bus stops, or whether two charging stations can be visited consecutively or not. Some \glspl{abk:vsp} rely on a discretized state-of-charge (cf.~\citealt{vanKootenNiekerk2017a}) or assume a given recharging time (cf.~\citealt{Li2014a}) to cope with the high planning complexity. 

Table~\ref{tab:vsp} shows to which extend existing \glspl{abk:vsp} fulfill the different practitioners' requirements.
By definition, \glspl{abk:vsp} assign trips to buses~(S1) and thereby also sequences to buses~(S2). As expected, most \glspl{abk:vsp} cover requirements for deadheading and detours~(D), while omitting requirements of charging infrastructure~(I). Moreover, \glspl{abk:vsp} rarely consider requirements of other categories. Remarkably, this also holds for requirements related to charging procedures~(C). Costs for buses~(T1) and energy~(T2) are covered by most \glspl{abk:vsp}, as at least one of these costs results from vehicle scheduling.
\begin{table*}[!hb]
	\centering
	\begin{threeparttable}
		\centering
		\newcommand{\PreserveBackslash}[1]{\let\temp=\\#1\let\\=\temp}
		\newcolumntype{C}[1]{>{\PreserveBackslash\centering}p{#1}}
		\setlength{\abovecaptionskip}{0.5ex}
		\caption{Fulfillment of the practitioners' requirements by \glspl{abk:clsap}.}
		\label{tab:clsap}
		\footnotesize
		\def\c{\checkmark}
		\addtolength{\tabcolsep}{-3.5pt}
		\singlespacing
		\begin{tabular}{l|l|llllll|lllll|llll|lll|ll|ll|ll}
			\toprule
			&& \multicolumn{6}{c|}{\textbf{T}} & \multicolumn{5}{c|}{\textbf{F}} & \multicolumn{4}{c|}{\textbf{I}}
			& \multicolumn{3}{c|}{\textbf{E}} & \multicolumn{2}{c|}{\textbf{C}} & \multicolumn{2}{c|}{\textbf{S}} & \multicolumn{2}{c}{\textbf{D}}\\
			& charg. loc. & 1 & 2 & 3 & 4 & 5 & 6 &
			1 & 2 & 3 & 4 & 5 &
			1 & 2 & 3 & 4 &
			1 & 2 & 3 &
			1 & 2 & 1 & 2 & 1 & 2\\
			\midrule
			\citealt{Jefferies2020a} &terminals&\c&\c&\c&\c&\c&\c& & &\c&\c&\c& &\c& &\c&\c&\c& & & & &\c&\c& \\
\citealt{Li2018c} &stations&\c&\c&\c&\c&\c&\c& & &\c&\c&\c& &\c& & & &\c& & &\c& &\c&\c&\c\\
\citealt{Li2019a} &stations&\c&\c&\c&\c&\c&\c& & &\c&\c&\c& &\c& & & &\c& & &\c& &\c&\c&\c\\
\citealt{Wei2018a} &terminals&\c& &\c& & & & & &\c& & & &\c& &\c& & & & & & &\c&\c&\c

			\\\bottomrule		
		\end{tabular}				
		\begin{tablenotes}				
			\scriptsize
			\item T: total costs, F: bus fleet, I: charging infrastructure, E: energy consumption, C: charging procedures, S: vehicle scheduling, D: deadheading and detours.
		\end{tablenotes}
		\centering
		\vspace{-0.3cm}
	\end{threeparttable}
\end{table*}

\section{Analysis and Perspectives}\label{sect:gaps}
In this section, we syntethize the analyses of Section~\ref{sect:literature}, discuss reasons for the identified gaps, and outline resulting research perspectives.

\begin{figure}[!hb]
	\centering
	\pgfplotstableread[col sep = semicolon]{text/pgf/figMethodClasses.csv}\DataForBars 
	\begin{tikzpicture}[scale=1.0]
	\begin{axis}[
	width=120mm,
	height=60mm,
	xlabel near ticks,
	ylabel near ticks,
	tick style = {black},
	enlarge x limits={abs=0.5},
	enlarge y limits=false,
	xtick pos=left,
	ytick pos=left,
	ybar stacked,
	ymin=0,
	ymax=20,
	xtick distance=1,
	minor x tick num=0,
	bar width=0.75cm,
	ylabel style={align=center},
	ylabel={Number of \\ publications},
	xticklabels={,,
		FTP,
		CLP,
		CLSAP,
		SAP,
		VSP,
	},
	]
	\addplot[fill=rwth,ultra thin] table [x=class,y=commercial] {\DataForBars};
	\addplot[fill=rwth-50,ultra thin] table [x=class,y=heuristic] {\DataForBars};
	\addplot[fill=rwth-10,ultra thin] table [x=class,y=exact] {\DataForBars};
	\end{axis}
	\begin{axis}[
	hide axis,
	width=120mm,
	height=60mm,
	xmin=0,
	xmax=1,
	ymin=0,
	ymax=1,
	legend style={at={(0.4,1)},anchor=north west,legend columns=1},
	legend cell align={left},
	]
	\addlegendimage{fill=rwth,mark=square*,only marks, mark size = 3.0, draw = black, ultra thin}
	\addlegendentry{commercial solver}
	\addlegendimage{fill=rwth-50,mark=square*,only marks, mark size = 3.0, draw = black, ultra thin}
	\addlegendentry{heuristic algorithm}
	\addlegendimage{fill=rwth-10,mark=square*,only marks, mark size = 3.0, draw = black, ultra thin}
	\addlegendentry{exact algorithm}
	\end{axis}
	\end{tikzpicture}
	\captionsetup{justification=centering}
	\caption{Applied solution methods per problem class.}
	\label{fig:methodClasses}
	\vspace{-0.7cm}
\end{figure}
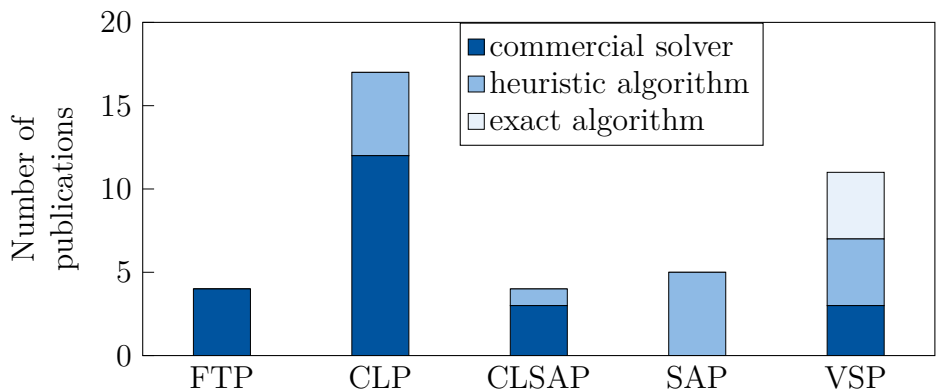

Planning the integration of \glspl{abk:beb} remains a complex planning task that may result in computationally intractable optimization problems. When applying such models to large-size instances of real-world bus networks, straightforwardly using off-the-shelf optimization software may find its limits depending on the problem type. Here, tailored heuristic or exact algorithms might be necessary to obtain reasonable solutions. Figure~\ref{fig:methodClasses} shows the share of publications that use commercial solvers or develop heuristic or exact algorithms for each problem class. As can be seen, mostly commercial solvers are used to solve planning problems in the more strategic problem classes ~(\glspl{abk:ftp}, \glspl{abk:clp}, \glspl{abk:clsap}). Contrary, commercial solvers are only rarely used to solve planning problems in the more operational problem classes~(\glspl{abk:sap}, \glspl{abk:vsp}). Here, exact algorithms based on branch-and-price and heuristic algorithms, often based on genetic or greedy algorithms, exist. While most problem classes have high combinatorial complexity, finding a solution to a strategic problem remains less time-critical than finding a solution to an operational problem, which partly explains the shift of solution approaches between the different problem classes. 
\begin{table*}[hb!]
	\centering
	\begin{threeparttable}
		\centering
		\newcommand{\PreserveBackslash}[1]{\let\temp=\\#1\let\\=\temp}
		\newcolumntype{C}[1]{>{\PreserveBackslash\centering}p{#1}}
		\setlength{\abovecaptionskip}{0.5ex}
		\caption{Fulfillment of the practitioners' requirements by \glspl{abk:vsp}.}
		\label{tab:vsp}
		\footnotesize
		\def\c{\checkmark}
		\addtolength{\tabcolsep}{-5pt}
		\singlespacing
		\begin{tabular}{l|l|llllll|lllll|llll|lll|ll|ll|ll}
			\toprule
			&& \multicolumn{6}{c|}{\textbf{T}} & \multicolumn{5}{c|}{\textbf{F}} & \multicolumn{4}{c|}{\textbf{I}}
			& \multicolumn{3}{c|}{\textbf{E}} & \multicolumn{2}{c|}{\textbf{C}} & \multicolumn{2}{c|}{\textbf{S}} & \multicolumn{2}{c}{\textbf{D}}\\
			& charg. loc. & 1 & 2 & 3 & 4 & 5 & 6 &
			1 & 2 & 3 & 4 & 5 &
			1 & 2 & 3 & 4 &
			1 & 2 & 3 &
			1 & 2 & 1 & 2 & 1 & 2\\
			\midrule
			\citealt{Adler2017a} &stations&\c& & & & &\c& & & & &\c& & & & & & & & & &\c&\c&\c&\c\\
\citealt{Janovec2019a} &terminals&\c&\c& & & & & & & & &\c& & & & & & & &\c& &\c&\c&\c&\c\\
\citealt{Li2014a} &swapping& & & &\c& &\c& & & & & & & & & & & & & & &\c&\c&\c&\c\\
\citealt{Liu2020a} &terminals&\c& &\c& & & & & & & & & & & & & & & &\c& &\c&\c&\c&\c\\
\citealt{Paul2014a} &terminals& & & & & &\c& & &\c& & & & & & & &\c& &\c& &\c&\c& & \\
\citealt{Rinaldi2019a} &terminals& & & & & &\c& & &\c& & & & & & & & & & & &\c&\c& & \\
\citealt{Rinaldi2019b} &terminals& & & & & &\c& & &\c& & & & & & & & & & & &\c&\c& & \\
\citealt{Tang2019a} &depot&\c& & & & &\c& & & & & & & & & &\c& &\c& & &\c&\c&\c&\c\\
\citealt{Teng2020a} &depot&\c& & & & &\c& & & &\c& & & & & & & & & & &\c&\c&\c&\c\\
\citealt{Wen2016a} &stations&\c& & & & &\c& & & & & & & & & & & & &\c& &\c&\c&\c&\c\\
\citealt{vanKootenNiekerk2017a} &stations&\c& & & & &\c& & & & & & & & & & & & &\c&\c&\c&\c&\c&\c

			\\\bottomrule
		\end{tabular}				
		\begin{tablenotes}				
			\scriptsize
			\item T: total costs, F: bus fleet, I: charging infrastructure, E: energy consumption, C: charging procedures, S: vehicle scheduling, D: deadheading and detours.
		\end{tablenotes}
		\centering
	\end{threeparttable}
	\vspace{-0.4cm}
\end{table*}

\begin{figure}[!hb]
	\centering
	\captionsetup{justification=centering}
	\begin{tikzpicture}
	\pgfplotstableread[col sep=semicolon]{text/pgf/figBoxPlot.csv}\csvdata
	\pgfplotstabletranspose\datatransposed{\csvdata} 
	\begin{axis}[
	width=120mm,
	height=60mm,
	xlabel near ticks,
	ylabel near ticks,
	tick style={black},
	enlarge x limits={abs=0.5},
	enlarge y limits=false,
	xtick pos=left,
	ytick pos=left,
	boxplot/draw direction=y,
	xtick = {1,2,3,4,5},
	xticklabels={
		FTP,
		CLP,
		CLSAP,
		SAP,
		VSP,
	},
	yticklabels={
		,0,20,40,60,
	},
	ylabel style={align=center},
	ylabel = {Coverage of practitioners'\\ requirements [\%]},
	ymin=0,ymax=.7,
	]
	\addplot+[boxplot,black] table[y index=1] {\datatransposed};
	\addplot+[boxplot,black] table[y index=2] {\datatransposed};
	\addplot+[boxplot,black] table[y index=3] {\datatransposed};
	\addplot+[boxplot,black] table[y index=4] {\datatransposed};
	\addplot+[boxplot,black] table[y index=5] {\datatransposed};
	\end{axis}
	\end{tikzpicture}	
	\caption{Average coverage of all practitioners' requirements per problem class.}
	\label{fig:boxPlot}
\end{figure}
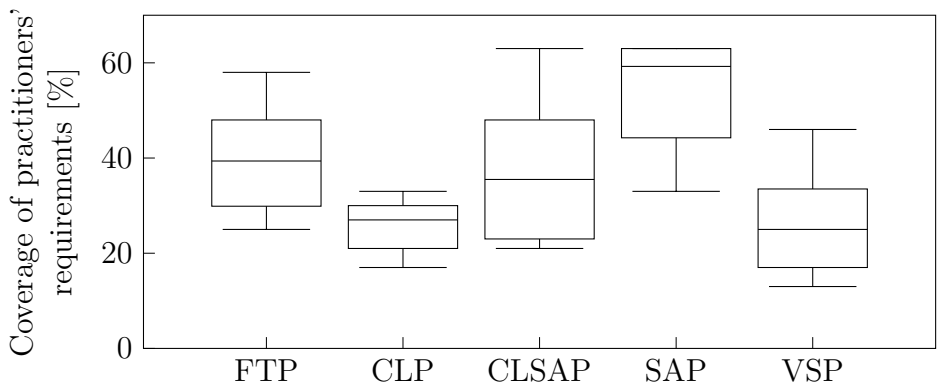

To further understand this shift of solution approaches between the different problem classes, Figure~\ref{fig:boxPlot} shows the share of practitioners' requirements covered by each publication and problem class as a Box-Whisker-Plot. On average, the publications achieve a relatively low coverage, indicating that the reviewed publications focus on specific requirements rather than on covering many. As can be seen, \glspl{abk:sap} and \glspl{abk:clsap} cover more requirements, while \glspl{abk:clp} and \glspl{abk:vsp} cover less requirements.  \glspl{abk:clp} usually do not account for operational constraints, which explains the low share of covered requirements linked to the high share of planning problems solved with commercial solvers. \glspl{abk:vsp} cover most operational requirements. However, solving such \glspl{abk:vsp} remains inherently hard such that although advanced solution methodologies are used, the overall share of covered requirements remains low to limit the computational complexity to its operational problem's core.  \glspl{abk:clsap} and \glspl{abk:sap} show a rather high share of covered requirements. For \glspl{abk:sap}, heuristic solution approaches are necessary to reach this share. For \glspl{abk:clsap}, such a heuristic sequence assignment is often included as an exogenous input, which then allows using commercial solvers to obtain good solutions. 

Three out of four publications focus on real-world case studies, analyzing a specific bus network. Together with Figures~\ref{fig:methodClasses}~and~\ref{fig:boxPlot}, this suggests that often some practitioner's requirements are withdrawn to preserve computational tractability in order to analyze one specific aspect in electric bus network optimization. In this context, we also note that common benchmark instances only exist for \glspl{abk:vsp} (cf., e.g.,~\citealt{Wen2016a}, \citealt{Adler2017a}). This shortage of common benchmark instances and solution methods suggests that academics so far lack a common definition of the respective planning problems. Vice versa, this reflects the multitude of potentially case-dependent practitioner requirements.

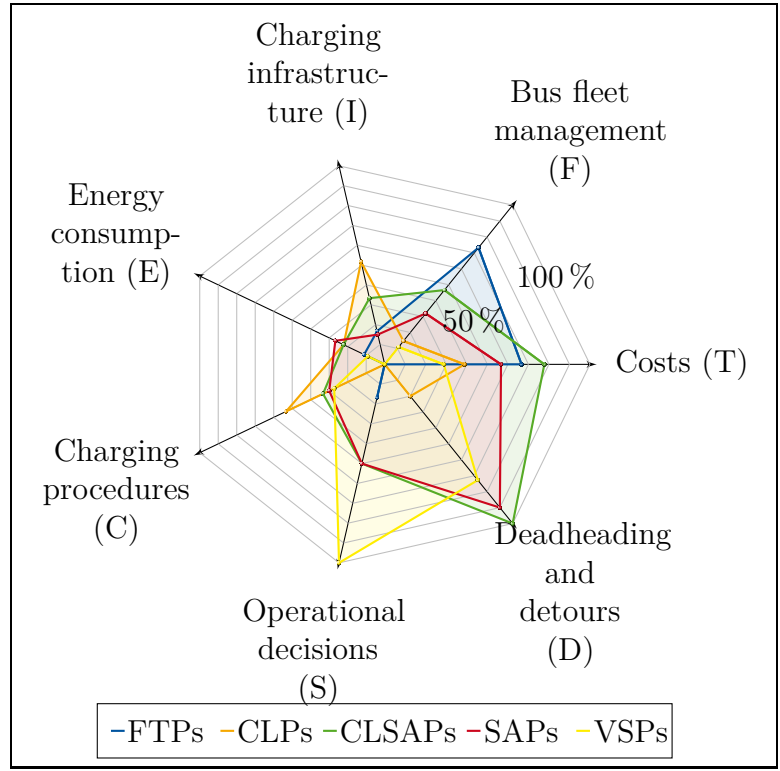
\begin{figure}[!ht]
	\centering
	\hspace*{1.275cm}
	\fbox{
		\begin{tikzpicture}
		\tkzKiviatDiagramFromFile[
		scale=.18,
		label distance=.5cm,
		gap     = 1.5,label space=6.75,  
		lattice = 10]{text/pgf/figSpider.dat}
		\tkzKiviatLineFromFile[thick,
		color      = rwth,
		mark       = ball,
		ball color = rwth,
		mark size  = 4pt,
		fill       = rwth!20]{text/pgf/figSpider.dat}{1}
		\tkzKiviatLineFromFile[thick,
		color      = orange,
		mark       = ball,
		ball color = orange,
		mark size  = 4pt,
		fill       = orange!20]{text/pgf/figSpider.dat}{2} 
		\tkzKiviatLineFromFile[thick,
		color      = grun,
		mark       = ball,
		ball color = grun,
		mark size  = 4pt,
		fill       = grun!20]{text/pgf/figSpider.dat}{3}
		\tkzKiviatLineFromFile[thick,
		color      = rot,
		mark       = ball,
		ball color = rot,
		mark size  = 4pt,
		fill       = rot!20]{text/pgf/figSpider.dat}{4}
		\tkzKiviatLineFromFile[thick,
		color      = yellow,
		mark       = ball,
		ball color = yellow,
		mark size  = 4pt,
		fill       = yellow!20]{text/pgf/figSpider.dat}{5}
		\draw (-21,-29) rectangle ++(42,4);
		\draw [thick, rwth] (-20,-27) -- (-19,-27); 
		\node at (-16,-27) {FTPs};
		\draw [thick, orange] (-12,-27) -- (-11,-27); 
		\node at (-8,-27) {CLPs};
		\draw [thick, grun] (-4.5,-27) -- (-3.5,-27); 
		\node at (1,-27) {CLSAPs};
		\draw [thick, rot] (6,-27) -- (7,-27); 
		\node at (10,-27) {SAPs};
		\draw [thick, yellow] (14,-27) -- (15,-27); 
		\node at (18,-27) {VSPs};
		\draw (12.5,6.5) node {100\,\%};
		\draw (6.5,3.25) node {50\,\%};
		\end{tikzpicture}}
	\captionsetup{justification=centering}
	\caption{Average coverage of practitioners' requirements per category and problem class.}
	\label{fig:spider}
\end{figure}

Figure~\ref{fig:spider} shows the share of requirements covered by publications of different problem classes in a Spider-Plot. To keep the figure concise, we aggregated the practitioners' requirements to the categories introduced in Section~\ref{sect:requirements} (cf.~Table~\ref{tab:requirements}). Figure~\ref{fig:spider} reveals gaps in all categories, as no problem class covers all of the practitioners' requirements. Particularly, requirements for charging infrastructure planning~(I), energy consumption~(E), and charging procedures~(C) often remain uncovered. The reason for missing coverage in charging infrastructure planning~(I) requirements is twofold. First, most models neglect charging infrastructure planning and assume a given charging infrastructure instead. Second, most models that address charging infrastructure planning focus exclusively on charging station locations but neglect additional decisions, e.g., 
on heterogeneous charging power levels 
or the number of charging stations per location. The missing energy consumption~(E) requirements result because most publications assume a constant energy consumption. This constitutes an unnecessary shortcoming as the consideration of realistic energy consumption is relatively straightforward from a modeling perspective. Incorporating realistic energy consumption improves the reliability of findings and decision support significantly without worsening the underlying model's computational complexity.
In contrast, the consideration of partial and non-linear recharging~(C) results in a high modeling complexity; however, this complexity is manageable by using a discretized battery state of charge (cf.~\citealt{vanKootenNiekerk2017a}) or a stepwise linearized recharging function (cf.~\citealt{Kunith2013a}).
Almost only \glspl{abk:ftp} consider requirements in the category of bus fleet management~(F), indicating that most publications so far focused primarily on a static design of electric bus networks rather than on time-dependent cost-optimal bus fleet transformations.

\begin{figure}[!ht]
	\centering
	\pgfplotstableread[col sep = semicolon]{text/pgf/figLocations.csv}\DataForBars 
	\begin{tikzpicture}[scale=1.0]
	\begin{axis}[
	width=120mm,
	height=60mm,
	xlabel near ticks,
	ylabel near ticks,
	tick style = {black},
	enlarge x limits={abs=0.5},
	enlarge y limits=false,
	xtick pos=left,
	ytick pos=left,
	ybar,
	xtick distance=1,
	bar width=0.35cm,
	ylabel style={align=center},
	ylabel={Share of covered\\ practitioners' requirements [\%]},
	xlabel={FTPs, CLPs, CLSAPs},
	every axis x label/.style={
		at={(ticklabel* cs:0.08,65)},
		anchor=west,
	},
	ymin=0, ymax=15,
	xticklabels={,,
		depot,
		terminals,
		stops,
		ex-route,
		swapping,
		depot,
		terminals,
		ex-route,
		swapping,
	},
	yticklabels={
		,0,20,40,60,
	},
	xticklabel style={rotate=45, anchor=north east},
	ymin=0, ymax=0.7,
	]
	\addplot[fill=rwth,ultra thin] table [x=class,y=avg] {\DataForBars};
	\node at (axis cs:3.5,4.5) [circle,draw,fill=white] {T};
	\path (axis cs:5,0) -- coordinate (m) (axis cs:6,0);
	\draw [dashed] (m) -- (current axis.north -| m);
	\end{axis}
	\begin{axis}[
	ticks=none,
	width=120mm,
	height=60mm,
	enlarge x limits={abs=0.5},
	enlarge y limits=false,
	xtick style = {black},
	xlabel={SAPs, VSPs},
	every axis x label/.style={
		at={(ticklabel* cs:0.675,65)},
		anchor=west,
	},
	hide y axis,
	ymin=0, ymax=0.7,
	legend style={at={(1,1)},anchor=north east},
	legend cell align={left},
	]
	\addplot[only marks] table [x=class,y=max] {\DataForBars};
	\addlegendimage{fill=rwth,mark=square*,only marks, mark size = 3.0, draw = black, ultra thin}
	\addlegendentry{max. coverage}                        
	\addlegendimage{fill=rwth-50,mark=square*,only marks, mark size = 3.0, draw = black, ultra thin}
	\addlegendentry{avg. coverage}
	\end{axis}
	\end{tikzpicture}
	\captionsetup{justification=centering}
	\caption{Average and maximum coverage of requirements per charging location.}
	\label{fig:location}
\end{figure}
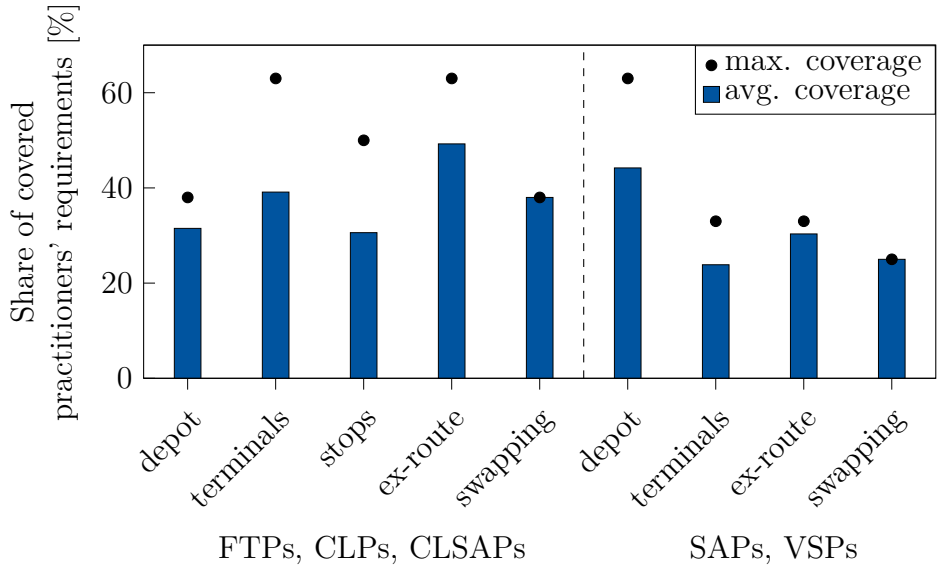

Furthermore, the coverage of practitioner's requirements varies depending on the charging station concept that a publication considers. Figure~\ref{fig:location} shows the average and maximum coverage for each class of charging station concept for strategically-focused~(\glspl{abk:ftp}, \glspl{abk:clp}, \glspl{abk:clsap}) and operationally-focused~(\glspl{abk:sap}, \glspl{abk:vsp}) publications.
Naturally, the considered charging station concept influences the coverage of requirements, as it affects the model complexity and thus the number of requirements that a model can cover from an overall computational complexity perspective.
Among publications on strategic planning, models that focus on an ex-route charging concept, i.e., charging at charging stations not located at bus network stops or depots, cover the largest share of practitioners' requirements. Here, a central assumption on clustering charging demand to high-capacity charging parks eases the integration of additional constraints and requirements. Contrarily, publications that focus on depot charging neglect several requirements.
In contrast, publications that focus on depot charging from an operational perspective cover more practitioners' requirements. Here, the simple concept of depot charging allows covering various constraints and requirements at an operational level. None of the operationally-focused publications considers recharging at in-line bus stops.

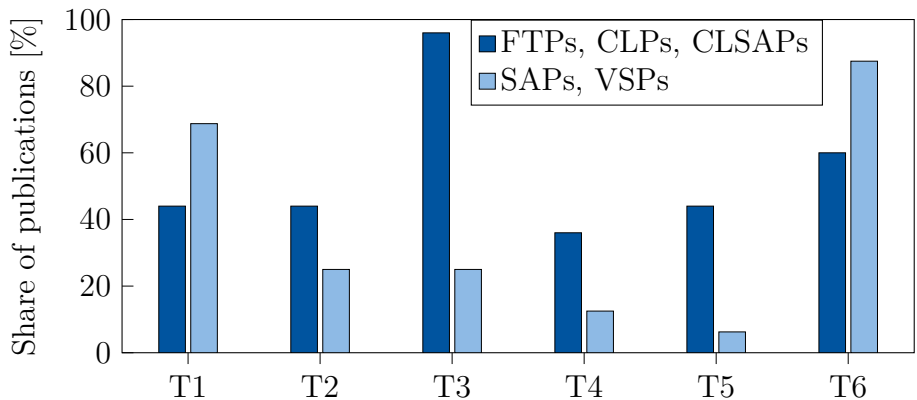
\begin{figure}[!hb]
	\centering
	\pgfplotstableread[col sep = semicolon]{text/pgf/figObj.csv}\DataForBars 
	\begin{tikzpicture}[scale=1.0]
	\begin{axis}[
	width=120mm,
	height=60mm,
	xlabel near ticks,
	ylabel near ticks,
	tick style = {black},
	enlarge x limits={abs=0.5},
	enlarge y limits=false,
	xtick pos=left,
	ytick pos=left,
	ybar,
	xtick distance=1,
	bar width=0.35cm,
	ymin=0,ymax=1,
	ylabel={Share of publications [\%]},
	yticklabels={
		,0,20,40,60,80,100
	},
	xticklabels={,,
		T1,
		T2,
		T3,
		T4,
		T5,
		T6,
	},
	]
	\addplot[fill=rwth,ultra thin] table [x=obj,y=strategic] {\DataForBars};
	\addplot[fill=rwth-50,ultra thin] table [x=obj,y=operational] {\DataForBars};
	\end{axis}
	\begin{axis}[
	hide axis,
	width=120mm,
	height=60mm,
	xmin=0,
	xmax=1,
	ymin=0,
	ymax=16,
	legend style={at={(0.44,1)},anchor=north west},
	legend cell align={left},
	]
	\addlegendimage{fill=rwth,mark=square*,only marks, mark size = 3.0, draw = black, ultra thin}
	\addlegendentry{FTPs, CLPs, CLSAPs}
	\addlegendimage{fill=rwth-50,mark=square*,only marks, mark size = 3.0, draw = black, ultra thin}
	\addlegendentry{SAPs, VSPs}
	\end{axis}
	\end{tikzpicture}
	\captionsetup{justification=centering}
	\caption{Coverage of total cost requirements in strategic and operational models.}
	\label{fig:objective}
\end{figure}

Figure~\ref{fig:objective} shows to which extent existing strategic (\glspl{abk:ftp}, \glspl{abk:clp}, \glspl{abk:clsap}) and operational (\glspl{abk:sap}, \glspl{abk:vsp}) problems consider relevant cost components~(T1-T6).
As can be seen, most strategic problem variants consider costs for charging stations and energy. The latter indicates that strategic problem variants aim to anticipate operations to a certain extend. Naturally, most operational problem variants consider energy costs. Some also cover rather strategic cost-components such as bus costs, as scheduling operations is a prerequisite for determining an optimal number of buses.
Some publications utilize a pure operationally-focused methodology but also aim at deriving strategic insights, e.g., regarding the required number of charging stations at the depot (cf.~\citealt{Rogge2018a}). We observe that overall maintenance costs are rarely considered.

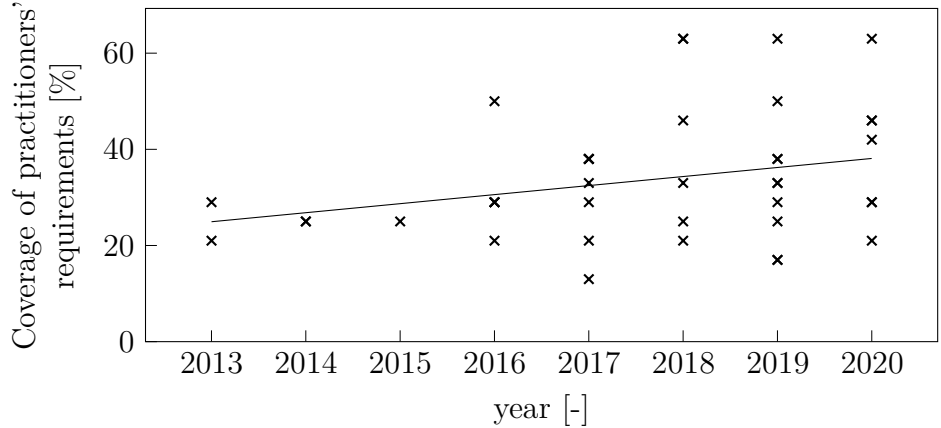
\begin{figure}[!ht]
	\centering
	\pgfplotstableread[col sep = semicolon]{text/pgf/figYearsData.csv}\DataA 
	\begin{tikzpicture}
	\begin{axis}[
	width=120mm,
	height=60mm,
	xlabel near ticks,
	ylabel near ticks,
	legend style={at={(0,1)},anchor=north west,
		/tikz/every odd column/.style={yshift=0pt},	
	},
	legend cell align={left},
	ytick pos=left,
	ylabel near ticks,
	tick style = {black},
	axis on top=true,
	xtick pos=left,
	restrict y to domain=-100:100,
	ylabel style={align=center},
	ylabel = {Coverage of practitioners'\\ requirements [\%]},
	xticklabels={,,
		2013,
		2014,
		2015,
		2016,
		2017,
		2018,
		2019,
		2020,
	},
	yticklabels={
		,0,20,40,60,
	},
	xlabel={year [-]},
	ymin=0,
	]
	\addplot[domain=2013:2020]{0.018806818*x-37.60875};
	\addplot[mark=x,black,only marks,mark size=2.5,mark options = {thick}] table [x=year,y=coverage]  {\DataA};
	\end{axis}
	\end{tikzpicture}
	\captionsetup{justification=centering}
	\caption{Average coverage of model requirements between 2013 and 2020.}
	\label{fig:years}
\end{figure}

Figure~\ref{fig:years} shows the temporal course of the coverage of the practitioners' requirements between~2013 and~2020. 
As can be seen, 
the coverage of the requirements 
has increased during the last years. The trend indicates academics' ambition to strive for more prosperous modeling approaches closer to real-world settings. However, it also reveals that the improvement in terms of covered practitioner's requirements is insidious such that a lot of open research questions remain to develop a comprehensive decision support system.
\section{Conclusion}\label{sect:conclusion}
In this review, we outlined the gap between practitioner's requirements and existing quantitative modeling approaches regarding the integration of \glspl{abk:beb} into existing urban bus networks. Herein, we identified requirements relevant for practitioners based on specifications of state-of-the-art \glspl{abk:beb}, project reports, and expert knowledge. We clustered the practitioners' requirements in categories related to total costs, bus fleet management, charging infrastructure, energy consumption, charging procedures, scheduling, and deadheading and detours. We then identified appropriate problem categories, namely \glsentryfullpl{abk:ftp}, \glsentryfullpl{abk:clp}, \glsentryfullpl{abk:clsap}, \glsentryfullpl{abk:sap} and \glsentryfullpl{abk:vsp}, and analyzed existing publications within these with regard to methodological aspects and covered practitioners' requirements. We found that the analyzed publications cover only a subset of practitioner's requirements, indicating a gap between academic research and practice. Only a few recent publications started to close this gap and cover some more requirements such that a variety of practitioners' requirements remain still uncovered. 

Particularly, the following main research gaps remain. First, some straightforward modeling extensions exist to sharpen the accuracy of planning approaches with respect to their real-world validity. Particularly, realistic energy consumption and recharging profiles can be incorporated without loosing computational tractability when using reasonable approximations, e.g., discretized battery state of charge or stepwise-linearized recharging functions. Moreover, additional cost components, e.g., maintenance cost can be added straightforwardly. Second, integrated modeling approaches that incorporate operational implications in strategic planning decisions promise to yield better planning solutions that meet practitioners' requirements. To this end, it remains an open question at which level of detail good operational surrogates can be found. First recent publications indicate that incorporating a sequence assignment decision constitutes an appropriate level of detail. Third, there exists a need for enhanced algorithmic solution techniques, which would ease the integration of additional constraints and would improve scalability of planning approaches towards large-scale instances. This is particularly the case for integrated modeling approaches that bear the highest computational complexity.

\section*{Acknowledgements}
Nicolas Dirks was funded by the PhD program “Access” by the German North-Rhine-Westphalian Ministry of Culture and Science in the framework of the funding scheme ”Forschungskollegs” (Grant no. 321-8.03.07-127598).

\footnotesize
\bibliographystyle{model5-names}
\bibliography{library}

\begin{thebibliography}{107}
\expandafter\ifx\csname natexlab\endcsname\relax\def\natexlab#1{#1}\fi
\providecommand{\url}[1]{\texttt{#1}}
\providecommand{\href}[2]{#2}
\providecommand{\path}[1]{#1}
\providecommand{\DOIprefix}{doi:}
\providecommand{\ArXivprefix}{arXiv:}
\providecommand{\URLprefix}{URL: }
\providecommand{\Pubmedprefix}{pmid:}
\providecommand{\doi}[1]{\href{http://dx.doi.org/#1}{\path{#1}}}
\providecommand{\Pubmed}[1]{\href{pmid:#1}{\path{#1}}}
\providecommand{\bibinfo}[2]{#2}
\ifx\xfnm\relax \def\xfnm[#1]{\unskip,\space#1}\fi
\bibitem[{ABB(2020{\natexlab{a}})}]{ABB2020b}
\bibinfo{author}{ABB} (\bibinfo{year}{2020}{\natexlab{a}}).
\newblock \bibinfo{title}{{ABB Ladesysteme - Beschreibung der Ladesäulen und
  ergänzender Dienstleistungen, Preislisten}}.
\newblock \URLprefix \url{https://new.abb.com/ev-charging/de}
  \bibinfo{note}{[last accessed: 02.04.2021]}.
\bibitem[{ABB(2020{\natexlab{b}})}]{ABB2020c}
\bibinfo{author}{ABB} (\bibinfo{year}{2020}{\natexlab{b}}).
\newblock \bibinfo{title}{Electric vehicle charging solutions}.
\newblock \URLprefix \url{https://new.abb.com/ev-charging/}
  \bibinfo{note}{[last accessed: 02.04.2021]}.
\bibitem[{ABB(2020{\natexlab{c}})}]{ABB2020a}
\bibinfo{author}{ABB} (\bibinfo{year}{2020}{\natexlab{c}}).
\newblock \bibinfo{title}{Pantograph down for electric buses}.
\newblock \URLprefix
  \url{https://new.abb.com/ev-charging/products/pantograph-down}
  \bibinfo{note}{[last accessed: 02.04.2021]}.
\bibitem[{Aber(2016)}]{Aber2016a}
\bibinfo{author}{Aber, J.} (\bibinfo{year}{2016}).
\newblock \bibinfo{title}{{Electric Bus Analysis for New York City Transit}}.
\newblock \bibinfo{institution}{Columbia University}.
\bibitem[{Adler \& Mirchandani(2017)}]{Adler2017a}
\bibinfo{author}{Adler, J.~D.}, \& \bibinfo{author}{Mirchandani, P.~B.}
  (\bibinfo{year}{2017}).
\newblock \bibinfo{title}{{The Vehicle Scheduling Problem for Fleets with
  Alternative-Fuel Vehicles}}.
\newblock {\it \bibinfo{journal}{Transportation Science}\/},  {\it
  \bibinfo{volume}{51}\/}, \bibinfo{pages}{441--456}.
  \DOIprefix\doi{10.1287/trsc.2015.0615}.
\bibitem[{AICTS(2018)}]{Atal2018a}
\bibinfo{author}{AICTS} (\bibinfo{year}{2018}).
\newblock \bibinfo{title}{{Purchase of Electric Vehicle (AC E- buses) from
  reputed electric vehicle manufacturing company under FAME scheme of Govt. of
  India}}.
\newblock \bibinfo{institution}{Atal Indore City Transport Services Limited}.
\bibitem[{Akasol(2020)}]{Akasol2020a}
\bibinfo{author}{Akasol} (\bibinfo{year}{2020}).
\newblock \bibinfo{title}{Akasystem 15 {OEM} 37 {PRC}}.
\newblock \URLprefix \url{https://www.akasol.com/de/akasystem-oem-prc}
  \bibinfo{note}{[last accessed: 02.04.2021]}.
\bibitem[{Akiyama et~al.(2001)Akiyama, Wahira, Kamata \& Fujii}]{Akiyama2001a}
\bibinfo{author}{Akiyama, T.}, \bibinfo{author}{Wahira, Y.},
  \bibinfo{author}{Kamata, M.}, \& \bibinfo{author}{Fujii, N.}
  (\bibinfo{year}{2001}).
\newblock \bibinfo{title}{Vehicle accessibility in japan today and the outlook
  for the future}.
\newblock {\it \bibinfo{journal}{{IATSS} Research}\/},  {\it
  \bibinfo{volume}{25}\/}, \bibinfo{pages}{42--50}.
  \DOIprefix\doi{10.1016/s0386-1112(14)60005-4}.
\newblock \bibinfo{note}{[last accessed: 02.04.2021]}.
\bibitem[{An(2020)}]{An2020a}
\bibinfo{author}{An, K.} (\bibinfo{year}{2020}).
\newblock \bibinfo{title}{{Battery electric bus infrastructure planning under
  demand uncertainty}}.
\newblock {\it \bibinfo{journal}{Transportation Research Part C: Emerging
  Technologies}\/},  {\it \bibinfo{volume}{111}\/}, \bibinfo{pages}{572--587}.
  \DOIprefix\doi{10.1016/j.trc.2020.01.009}.
\bibitem[{An et~al.(2019)An, Jing \& Kim}]{An2019a}
\bibinfo{author}{An, K.}, \bibinfo{author}{Jing, W.}, \& \bibinfo{author}{Kim,
  I.} (\bibinfo{year}{2019}).
\newblock \bibinfo{title}{Battery-swapping facility planning for electric buses
  with local charging systems}.
\newblock {\it \bibinfo{journal}{International Journal of Sustainable
  Transportation}\/},  {\it \bibinfo{volume}{14}\/}, \bibinfo{pages}{489--502}.
  \DOIprefix\doi{10.1080/15568318.2019.1573939}.
\bibitem[{{Astana LRT}(2018)}]{Nur-Sultan2018a}
\bibinfo{author}{{Astana LRT}} (\bibinfo{year}{2018}).
\newblock \bibinfo{title}{{Briefing on the topic: Development of public
  transport in Astana.}}
\newblock \URLprefix \url{http://www.alrt.kz/news/319} \bibinfo{note}{[last
  accessed: 02.04.2021]}.
\bibitem[{AVTA(2020)}]{AVTA2020a}
\bibinfo{author}{AVTA} (\bibinfo{year}{2020}).
\newblock \bibinfo{title}{About our fleet}.
\newblock \bibinfo{institution}{Antelope Valley Transit Authority}. \URLprefix
  \url{https://www.avta.com/about-our-fleet.php} \bibinfo{note}{[last accessed:
  02.04.2021]}.
\bibitem[{Berthold et~al.(2017)Berthold, Förster \& Rohrbeck}]{Berthold2017a}
\bibinfo{author}{Berthold, K.}, \bibinfo{author}{Förster, P.}, \&
  \bibinfo{author}{Rohrbeck, B.} (\bibinfo{year}{2017}).
\newblock \bibinfo{title}{{Location Planning of Charging Stations for Electric
  City Buses}}.
\newblock In {\it \bibinfo{booktitle}{Operations Research Proceedings}\/} (pp.
  \bibinfo{pages}{237--242}).
\newblock \DOIprefix\doi{10.1007/978-3-319-42902-1_32}.
\bibitem[{Bi et~al.(2018)Bi, Keoleian \& Ersal}]{Bi2018a}
\bibinfo{author}{Bi, Z.}, \bibinfo{author}{Keoleian, G.~A.}, \&
  \bibinfo{author}{Ersal, T.} (\bibinfo{year}{2018}).
\newblock \bibinfo{title}{{Wireless charger deployment for an electric bus
  network: A multi-objective life cycle optimization}}.
\newblock {\it \bibinfo{journal}{Applied Energy}\/},  {\it
  \bibinfo{volume}{225}\/}, \bibinfo{pages}{1090--1101}.
  \DOIprefix\doi{10.1016/j.apenergy.2018.05.070}.
\bibitem[{BlueSolutions(2019)}]{BlueSolutions2019a}
\bibinfo{author}{BlueSolutions} (\bibinfo{year}{2019}).
\newblock \bibinfo{title}{Pack {LMP} 63}.
\newblock \URLprefix
  \url{https://blue-storage.com/bollore-assets/uploads/2019/05/fiche-technique-pack-lmp-63.pdf}
  \bibinfo{note}{[last accessed: 02.04.2021]}.
\bibitem[{BVG(2018)}]{BVG2018a}
\bibinfo{author}{BVG} (\bibinfo{year}{2018}).
\newblock \bibinfo{title}{Zahlenspiegel 2019}.
\newblock \bibinfo{institution}{Public Transport Company of Berlin}. \URLprefix
  \url{https://unternehmen.bvg.de/de/Unternehmen/Medien/Publikationen}
  \bibinfo{note}{[last accessed: 02.04.2021]}.
\bibitem[{BVG(2020)}]{BVG2020a}
\bibinfo{author}{BVG} (\bibinfo{year}{2020}).
\newblock \bibinfo{title}{{E-Busse erobern Pankow und Lichtenberg}}.
\newblock \bibinfo{institution}{Public Transport Company of Berlin}. \URLprefix
  \url{https://www.bvg.de/de/Aktuell/Newsmeldung?newsid=3835}
  \bibinfo{note}{[last accessed: 02.04.2021]}.
\bibitem[{Chen \& Bierlaire(2013)}]{Chen2013a}
\bibinfo{author}{Chen, J.}, \& \bibinfo{author}{Bierlaire, M.}
  (\bibinfo{year}{2013}).
\newblock \bibinfo{title}{{Planning of feeding station installment for electric
  urban public mass-transportation system}}.
\newblock In {\it \bibinfo{booktitle}{13th Swiss Transportation Research
  Conference}\/} (pp. \bibinfo{pages}{1--12}).
\newblock \bibinfo{address}{Ascona, Switzerland}.
\newblock \URLprefix \url{https://core.ac.uk/download/pdf/148001806.pdf}
  \bibinfo{note}{[last accessed: 16.04.2021]}.
\bibitem[{Chen et~al.(2018)Chen, Yin \& Song}]{Chen2018a}
\bibinfo{author}{Chen, Z.}, \bibinfo{author}{Yin, Y.}, \&
  \bibinfo{author}{Song, Z.} (\bibinfo{year}{2018}).
\newblock \bibinfo{title}{{A cost-competitiveness analysis of charging
  infrastructure for electric bus operations}}.
\newblock {\it \bibinfo{journal}{Transportation Research Part C: Emerging
  Technologies}\/},  {\it \bibinfo{volume}{93}\/}, \bibinfo{pages}{351--366}.
  \DOIprefix\doi{10.1016/j.trc.2018.06.006}.
\bibitem[{Clairand et~al.(2019)Clairand, Guerra-Ter{\'{a}}n, Serrano-Guerrero,
  Gonz{\'{a}}lez-Rodr{\'{i}}guez \&
  Escriv{\'{a}}-Escriv{\'{a}}}]{Clairand2019a}
\bibinfo{author}{Clairand, J.-M.}, \bibinfo{author}{Guerra-Ter{\'{a}}n, P.},
  \bibinfo{author}{Serrano-Guerrero, X.},
  \bibinfo{author}{Gonz{\'{a}}lez-Rodr{\'{i}}guez, M.}, \&
  \bibinfo{author}{Escriv{\'{a}}-Escriv{\'{a}}, G.} (\bibinfo{year}{2019}).
\newblock \bibinfo{title}{{Electric Vehicles for Public Transportation in Power
  Systems: A Review of Methodologies}}.
\newblock {\it \bibinfo{journal}{Energies}\/},  {\it \bibinfo{volume}{12}\/},
  \bibinfo{pages}{1--22}. \DOIprefix\doi{10.3390/en12163114}.
\bibitem[{Daimler(2020)}]{Daimler2020a}
\bibinfo{author}{Daimler} (\bibinfo{year}{2020}).
\newblock \bibinfo{title}{omniplus services}.
\newblock \URLprefix \url{https://www.omniplus.com/uk} \bibinfo{note}{[last
  accessed: 02.04.2021]}.
\bibitem[{Denomme(2012)}]{Denomme2012a}
\bibinfo{author}{Denomme, G.} (\bibinfo{year}{2012}).
\newblock \bibinfo{title}{{Usable Energy: Key to Determining the True Cost of
  Advanced Lithium Ion Battery Systems for Electric Vehicles}}.
\newblock \URLprefix
  \url{https://www.batterypoweronline.com/markets/batteries/usable-energy-key-to-determining-the-true-cost-of-advanced-lithium-ion-battery-systems-for-electric-vehicles/}
  \bibinfo{note}{[last accessed: 02.04.2021]}.
\bibitem[{Dirks et~al.(2021)Dirks, Schiffer \& Walther}]{Dirks2021a}
\bibinfo{author}{Dirks, N.}, \bibinfo{author}{Schiffer, M.}, \&
  \bibinfo{author}{Walther, G.} (\bibinfo{year}{2021}).
\newblock \bibinfo{title}{On the integration of battery electric buses into
  urban bus networks}.
\newblock {\it \bibinfo{journal}{arXiv preprint arXiv:2103.12189}\/}.
\bibitem[{EAFO(2020)}]{EAFO2020a}
\bibinfo{author}{EAFO} (\bibinfo{year}{2020}).
\newblock \bibinfo{title}{Alternative fuel fleet (electricity) (2020)}.
\newblock \bibinfo{institution}{European Alternative Fuel Observatory}.
  \URLprefix \url{https://www.eafo.eu/vehicles-and-fleet/m2-m3}
  \bibinfo{note}{[last accessed: 02.04.2021]}.
\bibitem[{Ebusco(2020)}]{Ebusco2020a}
\bibinfo{author}{Ebusco} (\bibinfo{year}{2020}).
\newblock \bibinfo{title}{Ebusco 2.2 - made to move people}.
\newblock \URLprefix
  \url{https://www.ebusco.com/wp-content/uploads/EBUSCO_brochure_2-2_digi.pdf}
  \bibinfo{note}{[last accessed: 02.04.2021]}.
\bibitem[{Ekoenergetyka(2020)}]{Ekoenergetyka2020a}
\bibinfo{author}{Ekoenergetyka} (\bibinfo{year}{2020}).
\newblock \bibinfo{title}{Our product range}.
\newblock \URLprefix \url{https://ekoenergetyka.com.pl/charging-stations/}
  \bibinfo{note}{[last accessed: 02.04.2021]}.
\bibitem[{EvoBus(2020)}]{EvoBus2020a}
\bibinfo{author}{EvoBus} (\bibinfo{year}{2020}).
\newblock \bibinfo{title}{Der neue e{Citaro} - technische informationen}.
\newblock \URLprefix
  \url{https://www.mercedes-benz-bus.com/de_DE/models/ecitaro/facts/technical-data.html}
  \bibinfo{note}{[last accessed: 02.04.2021]}.
\bibitem[{Floman(2019)}]{Optibus2019a}
\bibinfo{author}{Floman, R.} (\bibinfo{year}{2019}).
\newblock \bibinfo{title}{{Understanding the Algorithmic Challenge of EV
  Optimization}}.
\newblock \bibinfo{institution}{Public transportation and bus scheduling
  software}. \URLprefix
  \url{https://www.optibus.com/understanding-the-algorithmic-challenges-of-ev-optimization/}
  \bibinfo{note}{[last accessed: 02.04.2021]}.
\bibitem[{Franca(2018)}]{Franca2018a}
\bibinfo{author}{Franca, A.} (\bibinfo{year}{2018}).
\newblock \bibinfo{title}{Electricity consumption and battery lifespan
  estimation for transit electric buses: drivetrain simulations and
  electrochemical modelling}.
\newblock \bibinfo{institution}{University of Victoria}.
\bibitem[{Glotz-Richter \& Koch(2018)}]{ELIPICT2018a}
\bibinfo{author}{Glotz-Richter, M.}, \& \bibinfo{author}{Koch, H.}
  (\bibinfo{year}{2018}).
\newblock \bibinfo{title}{{Use Cases - Electrification of public transport in
  cities}}.
\newblock \bibinfo{institution}{ELIPTIC}.
\bibitem[{Günter \& Koch(2018)}]{ELIPICTb}
\bibinfo{author}{Günter, H.}, \& \bibinfo{author}{Koch, H.}
  (\bibinfo{year}{2018}).
\newblock \bibinfo{title}{{Policy Recommendations - Electrification of public
  transport in cities}}.
\newblock \bibinfo{institution}{ELIPICT}.
\bibitem[{Guida \& Abdulah(2017)}]{ZeEUSProject2017a}
\bibinfo{author}{Guida, U.}, \& \bibinfo{author}{Abdulah, A.}
  (\bibinfo{year}{2017}).
\newblock \bibinfo{title}{{ZeEUS eBus Report No. 2. An updated overview of
  electric buses in Europe}}.
\newblock \URLprefix
  \url{http://zeeus.eu/uploads/publications/documents/zeeus-ebus-report-2.pdf}
  \bibinfo{note}{[last accessed: 16.04.2021]}.
\bibitem[{GVB(2020)}]{GVB2020a}
\bibinfo{author}{GVB} (\bibinfo{year}{2020}).
\newblock \bibinfo{title}{Mail [personal communication]}.
\newblock \bibinfo{institution}{Public Transport Company of Amsterdam}.
\bibitem[{{Hamburger Hochbahn}(2020)}]{Hochbahn2020a}
\bibinfo{author}{{Hamburger Hochbahn}} (\bibinfo{year}{2020}).
\newblock \bibinfo{title}{Mail [personal communication]}.
\newblock \bibinfo{institution}{Public Transport Company of Hamburg}.
\bibitem[{Hug(2015)}]{Hug2015a}
\bibinfo{author}{Hug, V.} (\bibinfo{year}{2015}).
\newblock \bibinfo{title}{Copenhagen trial with 12 m byd k9 electric buses}.
\newblock \bibinfo{institution}{Trafikselskabet Movia}.
\bibitem[{Impact(2018)}]{Impact2018a}
\bibinfo{author}{Impact} (\bibinfo{year}{2018}).
\newblock \bibinfo{title}{Uves {NMC} standard battery pack {NMC} 40 kwh}.
\newblock \URLprefix \url{https://icpt.pl/en/bus/} \bibinfo{note}{[last
  accessed: 02.04.2021]}.
\bibitem[{Init(2019)}]{Init2019a}
\bibinfo{author}{Init} (\bibinfo{year}{2019}).
\newblock \bibinfo{title}{The art of managing e-bus fleets}.
\newblock \bibinfo{institution}{Innovation in Traffic Systems SE}. \URLprefix
  \url{https://www.initse.com/fileadmin/user_upload/Content/3_Solutions/7_Electromobility/eMOBILE_system_brochure_en.pdf}
  \bibinfo{note}{[last accessed: 16.04.2021]}.
\bibitem[{{Intelligent Transport}(2019)}]{S.d.Chile2019b}
\bibinfo{author}{{Intelligent Transport}} (\bibinfo{year}{2019}).
\newblock \bibinfo{title}{{Latin America’s first 100 per cent electric bus
  corridor opens in Chile}}.
\newblock \URLprefix
  \url{https://www.intelligenttransport.com/transport-news/90769/latin-americas-first-100-per-cent-electric-bus-corridor-opens-in-chile/}
  \bibinfo{note}{[last accessed: 02.04.2021]}.
\bibitem[{Irizar(2020)}]{Irizar2020a}
\bibinfo{author}{Irizar} (\bibinfo{year}{2020}).
\newblock \bibinfo{title}{Comprehensive turnkey electromobility solutions for
  cities}.
\newblock \URLprefix
  \url{https://irizar.co.uk/wp-content/uploads/2019/11/catalogo_e-mobility_2019_en.pdf}
  \bibinfo{note}{[last accessed: 02.04.2021]}.
\bibitem[{Islam \& Lownes(2019)}]{Islam2019a}
\bibinfo{author}{Islam, A.}, \& \bibinfo{author}{Lownes, N.}
  (\bibinfo{year}{2019}).
\newblock \bibinfo{title}{{When to go electric? A parallel bus fleet
  replacement study}}.
\newblock {\it \bibinfo{journal}{Transportation Research Part D: Transport and
  Environment}\/},  {\it \bibinfo{volume}{72}\/}, \bibinfo{pages}{299--311}.
  \DOIprefix\doi{10.1016/j.trd.2019.05.007}.
\bibitem[{Janovec \& Koh{\'{a}}ni(2019)}]{Janovec2019a}
\bibinfo{author}{Janovec, M.}, \& \bibinfo{author}{Koh{\'{a}}ni, M.}
  (\bibinfo{year}{2019}).
\newblock \bibinfo{title}{{Exact approach to the electric bus fleet
  scheduling}}.
\newblock {\it \bibinfo{journal}{Transportation Research Procedia}\/},  {\it
  \bibinfo{volume}{40}\/}, \bibinfo{pages}{1380--1387}.
  \DOIprefix\doi{10.1016/j.trpro.2019.07.191}.
\bibitem[{Jefferies \& Göhlich(2020)}]{Jefferies2020a}
\bibinfo{author}{Jefferies, D.}, \& \bibinfo{author}{Göhlich, D.}
  (\bibinfo{year}{2020}).
\newblock \bibinfo{title}{A comprehensive {TCO} evaluation method for electric
  bus systems based on discrete-event simulation including bus scheduling and
  charging infrastructure optimisation}.
\newblock {\it \bibinfo{journal}{World Electric Vehicle Journal}\/},  {\it
  \bibinfo{volume}{11}\/}, \bibinfo{pages}{1--43}.
  \DOIprefix\doi{10.3390/wevj11030056}.
\bibitem[{Ke et~al.(2016)Ke, Chung \& Chen}]{Ke2016a}
\bibinfo{author}{Ke, B.-R.~R.}, \bibinfo{author}{Chung, C.-Y.~Y.}, \&
  \bibinfo{author}{Chen, Y.-C.~C.} (\bibinfo{year}{2016}).
\newblock \bibinfo{title}{{Minimizing the costs of constructing an all plug-in
  electric bus transportation system: A case study in Penghu}}.
\newblock {\it \bibinfo{journal}{Applied Energy}\/},  {\it
  \bibinfo{volume}{177}\/}, \bibinfo{pages}{649--660}.
  \DOIprefix\doi{10.1016/j.apenergy.2016.05.152}.
\bibitem[{Kliewer et~al.(2006)Kliewer, Mellouli \& Suhl}]{Kliewer2006a}
\bibinfo{author}{Kliewer, N.}, \bibinfo{author}{Mellouli, T.}, \&
  \bibinfo{author}{Suhl, L.} (\bibinfo{year}{2006}).
\newblock \bibinfo{title}{A time{\textendash}space network based exact
  optimization model for multi-depot bus scheduling}.
\newblock {\it \bibinfo{journal}{European Journal of Operational Research}\/},
  {\it \bibinfo{volume}{175}\/}, \bibinfo{pages}{1616--1627}.
  \DOIprefix\doi{10.1016/j.ejor.2005.02.030}.
\bibitem[{Koch(2017)}]{Koch2017a}
\bibinfo{author}{Koch, R.} (\bibinfo{year}{2017}).
\newblock \bibinfo{title}{{On-line Electrochemical Impedance - Spectroscopy for
  Lithium-Ion Battery Systems}}.
\newblock \bibinfo{institution}{Technical University of Munich}. \URLprefix
  \url{https://mediatum.ub.tum.de/doc/1343447/1343447.pdf} \bibinfo{note}{[last
  accessed: 02.04.2021]}.
\bibitem[{Kunith et~al.(2013)Kunith, Goehlich \& Mendelevitch}]{Kunith2013a}
\bibinfo{author}{Kunith, A.}, \bibinfo{author}{Goehlich, D.}, \&
  \bibinfo{author}{Mendelevitch, R.} (\bibinfo{year}{2013}).
\newblock \bibinfo{title}{{Planning and Optimization of Fast-Charging
  Infrastructure for Electric Urban Bus Systems}}.
\newblock In {\it \bibinfo{booktitle}{Proceedings of Second International
  Conference on Traffic and Transport Engineering (ICTTE)}\/} (pp.
  \bibinfo{pages}{43--49}).
\newblock \bibinfo{address}{Belgrade, Serbia}.
\bibitem[{Kunith et~al.(2017)Kunith, Mendelevitch \& Goehlich}]{Kunith2017b}
\bibinfo{author}{Kunith, A.}, \bibinfo{author}{Mendelevitch, R.}, \&
  \bibinfo{author}{Goehlich, D.} (\bibinfo{year}{2017}).
\newblock \bibinfo{title}{{Electrification of a city bus network - An
  optimization model for cost-effective placing of charging infrastructure and
  battery sizing of fast-charging electric bus systems}}.
\newblock {\it \bibinfo{journal}{International Journal of Sustainable
  Transportation}\/},  {\it \bibinfo{volume}{11}\/}, \bibinfo{pages}{707--720}.
  \DOIprefix\doi{10.1080/15568318.2017.1310962}.
\bibitem[{Kunith et~al.(2016)Kunith, Mendelevitch, Kuschmierz \&
  G{\"{o}}hlich}]{Kunith2016a}
\bibinfo{author}{Kunith, A.}, \bibinfo{author}{Mendelevitch, R.},
  \bibinfo{author}{Kuschmierz, A.}, \& \bibinfo{author}{G{\"{o}}hlich, D.}
  (\bibinfo{year}{2016}).
\newblock \bibinfo{title}{{Optimization of fast charging infrastructure for
  electric bus transportation - Electrification of a city bus network}}.
\newblock In {\it \bibinfo{booktitle}{EVS 2016 - 29th International Electric
  Vehicle Symposium}\/} (pp. \bibinfo{pages}{1--12}).
\newblock \bibinfo{address}{Montr{\'{e}}al, Qu{\'{e}}bec, Canada}.
\bibitem[{Lajunen(2018)}]{Lajunen2018a}
\bibinfo{author}{Lajunen, A.} (\bibinfo{year}{2018}).
\newblock \bibinfo{title}{{Lifecycle costs and charging requirements of
  electric buses with different charging methods}}.
\newblock {\it \bibinfo{journal}{Journal of Cleaner Production}\/},  {\it
  \bibinfo{volume}{172}\/}, \bibinfo{pages}{56--67}.
  \DOIprefix\doi{10.1016/j.jclepro.2017.10.066}.
\bibitem[{Li(2014)}]{Li2014a}
\bibinfo{author}{Li, J.-Q.} (\bibinfo{year}{2014}).
\newblock \bibinfo{title}{{Transit Bus Scheduling with Limited Energy}}.
\newblock {\it \bibinfo{journal}{Transportation Science}\/},  {\it
  \bibinfo{volume}{48}\/}, \bibinfo{pages}{521--539}.
  \DOIprefix\doi{10.1287/trsc.2013.0468}.
\bibitem[{Li et~al.(2018{\natexlab{a}})Li, Lo \& Xiao}]{Li2018c}
\bibinfo{author}{Li, L.}, \bibinfo{author}{Lo, H.~K.}, \&
  \bibinfo{author}{Xiao, F.} (\bibinfo{year}{2018}{\natexlab{a}}).
\newblock \bibinfo{title}{{Optimizing mixed-fleet bus scheduling under range
  constraint}}.
\newblock In {\it \bibinfo{booktitle}{CASPT 2018 (Conference on Advanced
  Systems in Public Transport and Transit Data 2018)}\/} (pp.
  \bibinfo{pages}{1--16}).
\newblock \bibinfo{address}{Brisbane, Australia}.
\newblock \URLprefix
  \url{http://www.caspt.org/wp-content/uploads/2018/10/Papers/CASPT_2018_paper_72.pdf}
  \bibinfo{note}{[last accessed: 16.04.201]}.
\bibitem[{Li et~al.(2019)Li, Lo \& Xiao}]{Li2019a}
\bibinfo{author}{Li, L.}, \bibinfo{author}{Lo, H.~K.}, \&
  \bibinfo{author}{Xiao, F.} (\bibinfo{year}{2019}).
\newblock \bibinfo{title}{{Mixed bus fleet scheduling under range and refueling
  constraints}}.
\newblock {\it \bibinfo{journal}{Transportation Research Part C: Emerging
  Technologies}\/},  {\it \bibinfo{volume}{104}\/}, \bibinfo{pages}{443--462}.
  \DOIprefix\doi{10.1016/j.trc.2019.05.009}.
\bibitem[{Li et~al.(2018{\natexlab{b}})Li, Lo, Xiao \& Cen}]{Li2018a}
\bibinfo{author}{Li, L.}, \bibinfo{author}{Lo, H.~K.}, \bibinfo{author}{Xiao,
  F.}, \& \bibinfo{author}{Cen, X.} (\bibinfo{year}{2018}{\natexlab{b}}).
\newblock \bibinfo{title}{{Mixed bus fleet management strategy for minimizing
  overall and emissions external costs}}.
\newblock {\it \bibinfo{journal}{Transportation Research Part D: Transport and
  Environment}\/},  {\it \bibinfo{volume}{60}\/}, \bibinfo{pages}{104--118}.
  \DOIprefix\doi{10.1016/j.trd.2016.10.001}.
\bibitem[{Liksutov(2020)}]{RussellPublishing2020a}
\bibinfo{author}{Liksutov, M.~S.} (\bibinfo{year}{2020}).
\newblock \bibinfo{title}{{City Snapshot: Mobility in Moscow}}.
\newblock \URLprefix
  \url{https://www.intelligenttransport.com/transport-articles/95256/city-snapshot-mobility-in-moscow/}
  \bibinfo{note}{[last accessed: 02.04.2021]}.
\bibitem[{Lin et~al.(2019{\natexlab{a}})Lin, Zhang, Shen \& Miao}]{Lin2019a}
\bibinfo{author}{Lin, Y.}, \bibinfo{author}{Zhang, K.}, \bibinfo{author}{Shen,
  Z.-J.~M.}, \& \bibinfo{author}{Miao, L.}
  (\bibinfo{year}{2019}{\natexlab{a}}).
\newblock \bibinfo{title}{{Charging Network Planning for Electric Bus Cities: A
  Case Study of Shenzhen, China}}.
\newblock {\it \bibinfo{journal}{Sustainability}\/},  {\it
  \bibinfo{volume}{11}\/}, \bibinfo{pages}{1--27}.
  \DOIprefix\doi{10.3390/su11174713}.
\bibitem[{Lin et~al.(2019{\natexlab{b}})Lin, Zhang, Shen, Ye \&
  Miao}]{Lin2019b}
\bibinfo{author}{Lin, Y.}, \bibinfo{author}{Zhang, K.}, \bibinfo{author}{Shen,
  Z. J.~M.}, \bibinfo{author}{Ye, B.}, \& \bibinfo{author}{Miao, L.}
  (\bibinfo{year}{2019}{\natexlab{b}}).
\newblock \bibinfo{title}{{Multistage large-scale charging station planning for
  electric buses considering transportation network and power grid}}.
\newblock {\it \bibinfo{journal}{Transportation Research Part C: Emerging
  Technologies}\/},  {\it \bibinfo{volume}{107}\/}, \bibinfo{pages}{423--443}.
  \DOIprefix\doi{10.1016/j.trc.2019.08.009}.
\bibitem[{Liu \& Ceder(2020)}]{Liu2020a}
\bibinfo{author}{Liu, T.}, \& \bibinfo{author}{Ceder, A.~A.}
  (\bibinfo{year}{2020}).
\newblock \bibinfo{title}{Battery-electric transit vehicle scheduling with
  optimal number of stationary chargers}.
\newblock {\it \bibinfo{journal}{Transportation Research Part C: Emerging
  Technologies}\/},  {\it \bibinfo{volume}{114}\/}, \bibinfo{pages}{118--139}.
  \DOIprefix\doi{10.1016/j.trc.2020.02.009}.
\bibitem[{Liu et~al.(2018)Liu, Song \& He}]{Liu2018b}
\bibinfo{author}{Liu, Z.}, \bibinfo{author}{Song, Z.}, \& \bibinfo{author}{He,
  Y.} (\bibinfo{year}{2018}).
\newblock \bibinfo{title}{{Planning of Fast-Charging Stations for a Battery
  Electric Bus System under Energy Consumption Uncertainty}}.
\newblock {\it \bibinfo{journal}{Transportation Research Record: Journal of the
  Transportation Research Board}\/},  {\it \bibinfo{volume}{2672}\/},
  \bibinfo{pages}{96--107}. \DOIprefix\doi{10.1177/0361198118772953}.
\bibitem[{Lotfi et~al.(2020)Lotfi, Pereira, Paterakis, Gabbar \&
  Catalao}]{Lotfi2020a}
\bibinfo{author}{Lotfi, M.}, \bibinfo{author}{Pereira, P.},
  \bibinfo{author}{Paterakis, N.~G.}, \bibinfo{author}{Gabbar, H.~A.}, \&
  \bibinfo{author}{Catalao, J. P.~S.} (\bibinfo{year}{2020}).
\newblock \bibinfo{title}{Optimal design of electric bus transport systems with
  minimal total ownership cost}.
\newblock {\it \bibinfo{journal}{{IEEE} Access}\/},  {\it
  \bibinfo{volume}{8}\/}, \bibinfo{pages}{119184--119199}.
  \DOIprefix\doi{10.1109/access.2020.3004910}.
\bibitem[{MAN(2020)}]{MAN2020a}
\bibinfo{author}{MAN} (\bibinfo{year}{2020}).
\newblock \bibinfo{title}{Einfach einsteigen. in die mobilität der zukunft.
  ihre mobilitätslösung von man.}
\newblock \URLprefix
  \url{https://www.truck.man.eu/man/media/de/content_medien/images/business_websites_suisse/broschueren_bilder_downloadbereich/bus_13/stadtbusse_2/8_br_man_emobility_broschuere_de_screen.pdf}
  \bibinfo{note}{[last accessed: 02.04.2021]}.
\bibitem[{Mathieu(2018)}]{Mathieu2018a}
\bibinfo{author}{Mathieu, L.} (\bibinfo{year}{2018}).
\newblock \bibinfo{title}{{Electric buses arrive on time - Marketplace,
  economic, technology, environmental and policy perspectives for fully
  electric buses in the EU}}.
\newblock \bibinfo{institution}{Transport {\&} Environment}. \URLprefix
  \url{https://www.transportenvironment.org/sites/te/files/Electric buses
  arrive on time.pdf} \bibinfo{note}{[last accessed: 16.04.2021]}.
\bibitem[{{Momentum Wireless Power}(2020)}]{Momentum2020}
\bibinfo{author}{{Momentum Wireless Power}} (\bibinfo{year}{2020}).
\newblock \bibinfo{title}{Wireless charging in action: Public transport}.
\newblock \URLprefix \url{https://momentumdynamics.com/solution/#publictransit}
  \bibinfo{note}{[last accessed: 02.04.2021]}.
\bibitem[{{Moscow Gov.}(2019)}]{Moscow2019a}
\bibinfo{author}{{Moscow Gov.}} (\bibinfo{year}{2019}).
\newblock \bibinfo{title}{{Electric bus. Everything you need to know about
  eco-friendly urban transport}}.
\newblock \URLprefix \url{https://www.mos.ru/en/news/item/64207073/}
  \bibinfo{note}{[last accessed: 02.04.2021]}.
\bibitem[{Naumann \& Vogelpohl(2015)}]{CACTUS2015a}
\bibinfo{author}{Naumann, S.}, \& \bibinfo{author}{Vogelpohl, H.}
  (\bibinfo{year}{2015}).
\newblock \bibinfo{title}{{Deliverable 1.2: Technologies for Fully Electric
  Busses}}.
\newblock \bibinfo{institution}{CACTUS - Electromobility}.
\bibitem[{Paul \& Yamada(2014)}]{Paul2014a}
\bibinfo{author}{Paul, T.}, \& \bibinfo{author}{Yamada, H.}
  (\bibinfo{year}{2014}).
\newblock \bibinfo{title}{{Operation and Charging Scheduling of Electric Buses
  in a City Bus Route Network}}.
\newblock In {\it \bibinfo{booktitle}{17th International IEEE Conference on
  Intelligent Transportation Systems (ITSC)}\/} (pp.
  \bibinfo{pages}{2780--2786}).
\newblock \bibinfo{publisher}{IEEE} volume~\bibinfo{volume}{1}.
\newblock \URLprefix \url{http://ieeexplore.ieee.org/document/6958135/}.
  \DOIprefix\doi{10.1109/ITSC.2014.6958135}.
\bibitem[{Pelletier et~al.(2017)Pelletier, Jabali, Laporte \&
  Veneroni}]{Pelletier2017a}
\bibinfo{author}{Pelletier, S.}, \bibinfo{author}{Jabali, O.},
  \bibinfo{author}{Laporte, G.}, \& \bibinfo{author}{Veneroni, M.}
  (\bibinfo{year}{2017}).
\newblock \bibinfo{title}{{Battery degradation and behaviour for electric
  vehicles: Review and numerical analyses of several models}}.
\newblock {\it \bibinfo{journal}{Transportation Research Part B:
  Methodological}\/},  {\it \bibinfo{volume}{103}\/},
  \bibinfo{pages}{158--187}. \DOIprefix\doi{10.1016/j.trb.2017.01.020}.
\bibitem[{Pelletier et~al.(2019)Pelletier, Jabali, Mendoza \&
  Laporte}]{Pelletier2019a}
\bibinfo{author}{Pelletier, S.}, \bibinfo{author}{Jabali, O.},
  \bibinfo{author}{Mendoza, J.~E.}, \& \bibinfo{author}{Laporte, G.}
  (\bibinfo{year}{2019}).
\newblock \bibinfo{title}{{The electric bus fleet transition problem}}.
\newblock {\it \bibinfo{journal}{Transportation Research Part C: Emerging
  Technologies}\/},  {\it \bibinfo{volume}{109}\/}, \bibinfo{pages}{174--193}.
  \DOIprefix\doi{10.1016/j.trc.2019.10.012}.
\bibitem[{{Phoenix Contact}(2017)}]{PhoenixContact2017a}
\bibinfo{author}{{Phoenix Contact}} (\bibinfo{year}{2017}).
\newblock \bibinfo{title}{High power charging - {CCS}-based fast charging with
  up to 500 {A}}.
\newblock \bibinfo{note}{[last accessed: 02.04.2021]}.
\bibitem[{Proterra(2020)}]{Proterra2020a}
\bibinfo{author}{Proterra} (\bibinfo{year}{2020}).
\newblock \bibinfo{title}{Energy fleet solutions}.
\newblock \bibinfo{institution}{Manufacturer of zero-emission buses}.
  \URLprefix \url{https://www.proterra.com/energy-services/}
  \bibinfo{note}{[last accessed: 02.04.2021]}.
\bibitem[{Red(2019)}]{S.d.Chile2019a}
\bibinfo{author}{Red} (\bibinfo{year}{2019}).
\newblock \bibinfo{title}{99 por ciento de la flota de buses está operativa
  este domingo y metro mantiene funcionamiento parcial de líneas 1, 2, 3, 5 y
  6}.
\newblock \bibinfo{institution}{Public Transport Company of Santiago de Chile}.
  \URLprefix
  \url{http://www.red.cl/noticias/99-de-la-flota-de-buses-esta-operativa-este-domingo-y-metro-mantiene-funcionamiento-parcial-de-lineas-1-2-3-5-y-6}
  \bibinfo{note}{[last accessed: 02.04.2021]}.
\bibitem[{Reuer et~al.(2015)Reuer, Kliewer \& Wolbeck}]{Reuer2015a}
\bibinfo{author}{Reuer, J.}, \bibinfo{author}{Kliewer, N.}, \&
  \bibinfo{author}{Wolbeck, L.} (\bibinfo{year}{2015}).
\newblock \bibinfo{title}{{The Electric Vehicle Scheduling Problem}}.
\newblock In {\it \bibinfo{booktitle}{CASPT 2015 (Conference on Advanced
  Systems in Public Transport 2015)}\/}.
\newblock \bibinfo{address}{Rotterdam, Netherlands}.
\newblock \URLprefix
  \url{http://www.rotterdam2015.caspt.org/proceedings/paper93.pdf}
  \bibinfo{note}{[last accessed: 16.04.2021]}.
\bibitem[{Rinaldi et~al.(2019{\natexlab{a}})Rinaldi, Picarelli, D'Ariano \&
  Viti}]{Rinaldi2019a}
\bibinfo{author}{Rinaldi, M.}, \bibinfo{author}{Picarelli, E.},
  \bibinfo{author}{D'Ariano, A.}, \& \bibinfo{author}{Viti, F.}
  (\bibinfo{year}{2019}{\natexlab{a}}).
\newblock \bibinfo{title}{{Mixed-fleet single-terminal bus scheduling problem:
  Modelling, solution scheme and potential applications}}.
\newblock {\it \bibinfo{journal}{Omega}\/},  {\it \bibinfo{volume}{96}\/},
  \bibinfo{pages}{1--39}. \DOIprefix\doi{10.1016/j.omega.2019.05.006}.
\bibitem[{Rinaldi et~al.(2019{\natexlab{b}})Rinaldi, Picarelli, Laskaris,
  D'Ariano \& Viti}]{Rinaldi2019b}
\bibinfo{author}{Rinaldi, M.}, \bibinfo{author}{Picarelli, E.},
  \bibinfo{author}{Laskaris, G.}, \bibinfo{author}{D'Ariano, A.}, \&
  \bibinfo{author}{Viti, F.} (\bibinfo{year}{2019}{\natexlab{b}}).
\newblock \bibinfo{title}{{Mixed hybrid and electric bus dynamic fleet
  management in urban networks: a model predictive control approach}}.
\newblock In {\it \bibinfo{booktitle}{2019 6th International Conference on
  Models and Technologies for Intelligent Transportation Systems (MT-ITS)}\/}
  (pp. \bibinfo{pages}{1--8}).
\newblock \bibinfo{publisher}{IEEE}.
\newblock \URLprefix \url{https://ieeexplore.ieee.org/document/8883387/}.
  \DOIprefix\doi{10.1109/MTITS.2019.8883387}.
\bibitem[{Rogge et~al.(2018)Rogge, van~der Hurk, Larsen \& Sauer}]{Rogge2018a}
\bibinfo{author}{Rogge, M.}, \bibinfo{author}{van~der Hurk, E.},
  \bibinfo{author}{Larsen, A.}, \& \bibinfo{author}{Sauer, D.~U.}
  (\bibinfo{year}{2018}).
\newblock \bibinfo{title}{{Electric bus fleet size and mix problem with
  optimization of charging infrastructure}}.
\newblock {\it \bibinfo{journal}{Applied Energy}\/},  {\it
  \bibinfo{volume}{211}\/}, \bibinfo{pages}{282--295}.
  \DOIprefix\doi{10.1016/j.apenergy.2017.11.051}.
\bibitem[{Rohrbeck et~al.(2018)Rohrbeck, Berthold \& Hettich}]{Rohrbeck2018a}
\bibinfo{author}{Rohrbeck, B.}, \bibinfo{author}{Berthold, K.}, \&
  \bibinfo{author}{Hettich, F.} (\bibinfo{year}{2018}).
\newblock \bibinfo{title}{{Location Planning of Charging Stations for Electric
  City Buses Considering Battery Ageing Effects}}.
\newblock In {\it \bibinfo{booktitle}{Operations Research Proceedings 2015}\/}
  (pp. \bibinfo{pages}{701--707}).
\newblock \DOIprefix\doi{10.1007/978-3-319-89920-6_93}.
\bibitem[{Sassi \& Oulamara(2017)}]{Sassi2017a}
\bibinfo{author}{Sassi, O.}, \& \bibinfo{author}{Oulamara, A.}
  (\bibinfo{year}{2017}).
\newblock \bibinfo{title}{{Electric vehicle scheduling and optimal charging
  problem: complexity, exact and heuristic approaches}}.
\newblock {\it \bibinfo{journal}{International Journal of Production
  Research}\/},  {\it \bibinfo{volume}{55}\/}, \bibinfo{pages}{519--535}.
  \DOIprefix\doi{10.1080/00207543.2016.1192695}.
\bibitem[{Scania(2020)}]{Scania2020a}
\bibinfo{author}{Scania} (\bibinfo{year}{2020}).
\newblock \bibinfo{title}{Fully electric low floor bus - scania citywide}.
\newblock \URLprefix
  \url{https://www.scania.com/content/dam/scanianoe/market/bg/products-and-services/buses-and-coaches/bev.pdf}
  \bibinfo{note}{[last accessed: 02.04.2021]}.
\bibitem[{Schiffer \& Walther(2017)}]{Schiffer2017a}
\bibinfo{author}{Schiffer, M.}, \& \bibinfo{author}{Walther, G.}
  (\bibinfo{year}{2017}).
\newblock \bibinfo{title}{The electric location routing problem with time
  windows and partial recharging}.
\newblock {\it \bibinfo{journal}{European Journal of Operational Research}\/},
  {\it \bibinfo{volume}{260}\/}, \bibinfo{pages}{995--1013}.
  \DOIprefix\doi{10.1016/j.ejor.2017.01.011}.
\bibitem[{Sebastiani et~al.(2016)Sebastiani, Luders \&
  Fonseca}]{Sebastiani2016a}
\bibinfo{author}{Sebastiani, M.~T.}, \bibinfo{author}{Luders, R.}, \&
  \bibinfo{author}{Fonseca, K.~V.} (\bibinfo{year}{2016}).
\newblock \bibinfo{title}{{Evaluating Electric Bus Operation for a Real-World
  BRT Public Transportation Using Simulation Optimization}}.
\newblock {\it \bibinfo{journal}{IEEE Transactions on Intelligent
  Transportation Systems}\/},  {\it \bibinfo{volume}{17}\/},
  \bibinfo{pages}{2777--2786}. \DOIprefix\doi{10.1109/TITS.2016.2525800}.
\bibitem[{Siemens(2020)}]{Siemens2020a}
\bibinfo{author}{Siemens} (\bibinfo{year}{2020}).
\newblock \bibinfo{title}{Charging systems for ebuses}.
\newblock \URLprefix
  \url{https://new.siemens.com/global/en/products/energy/medium-voltage/solutions/emobility/ebus-depot.html}
  \bibinfo{note}{[last accessed: 02.04.2021]}.
\bibitem[{Sinhuber et~al.(2012)Sinhuber, Rohlfs \& Sauer}]{Sinhuber2012a}
\bibinfo{author}{Sinhuber, P.}, \bibinfo{author}{Rohlfs, W.}, \&
  \bibinfo{author}{Sauer, D.~U.} (\bibinfo{year}{2012}).
\newblock \bibinfo{title}{{Study on Power and Energy Demand for Sizing the
  Energy Storage Systems for Electrified Local Public Transport Buses}}.
\newblock In {\it \bibinfo{booktitle}{2012 IEEE Vehicle Power and Propulsion
  Conference}\/} (pp. \bibinfo{pages}{315--320}).
\newblock \bibinfo{publisher}{IEEE}.
\newblock \DOIprefix\doi{10.1109/VPPC.2012.6422680}.
\bibitem[{{Stagecoach}(2020{\natexlab{a}})}]{Manchester2020b}
\bibinfo{author}{{Stagecoach}} (\bibinfo{year}{2020}{\natexlab{a}}).
\newblock \bibinfo{title}{{About Stagecoach Manchester - Our fleet}}.
\newblock \bibinfo{institution}{Public Transport Company of Greater
  Manchester}. \URLprefix
  \url{https://www.stagecoachbus.com/about/manchester#tab6}
  \bibinfo{note}{[last accessed: 02.04.2021]}.
\bibitem[{{Stagecoach}(2020{\natexlab{b}})}]{Manchester2020a}
\bibinfo{author}{{Stagecoach}} (\bibinfo{year}{2020}{\natexlab{b}}).
\newblock \bibinfo{title}{{First electric doubledecker buses for Greater
  Manchester}}.
\newblock \bibinfo{institution}{Public Transport Company of Greater
  Manchester}. \URLprefix
  \url{https://www.stagecoachbus.com/news/manchester/2020/march/first-double-deck-electric-buses-for-greater-manchester}
  \bibinfo{note}{[last accessed: 02.04.2021]}.
\bibitem[{{Sustainable Bus}(2020{\natexlab{a}})}]{Nur-Sultan2020a}
\bibinfo{author}{{Sustainable Bus}} (\bibinfo{year}{2020}{\natexlab{a}}).
\newblock \bibinfo{title}{{100 Yutong electric buses delivered in
  Kazakhstan’s capital Nur-Sultan}}.
\newblock \URLprefix
  \url{https://www.sustainable-bus.com/electric-bus/100-yutong-electric-buses-delivered-in-kazakhstans-capital-city-nur-sultan/}
  \bibinfo{note}{[last accessed: 02.04.2021]}.
\bibitem[{{Sustainable Bus}(2020{\natexlab{b}})}]{Madrid2020a}
\bibinfo{author}{{Sustainable Bus}} (\bibinfo{year}{2020}{\natexlab{b}}).
\newblock \bibinfo{title}{{50 electric buses in the future of Madrid. EMT to
  invest 35 million euros}}.
\newblock \URLprefix
  \url{https://www.sustainable-bus.com/electric-bus/50-electric-buses-in-the-future-of-madrid-emt-to-invest-35-million-euros/}
  \bibinfo{note}{[last accessed: 02.04.2021]}.
\bibitem[{{Sustainable Bus}(2020{\natexlab{c}})}]{Danish2020a}
\bibinfo{author}{{Sustainable Bus}} (\bibinfo{year}{2020}{\natexlab{c}}).
\newblock \bibinfo{title}{{The six largest Danish cities commit to buying only
  ZE buses from 2021 on}}.
\newblock \URLprefix
  \url{https://www.sustainable-bus.com/news/the-six-largest-danish-cities-commit-to-buying-only-ze-buses-from-2021-on/}
  \bibinfo{note}{[last accessed: 02.04.2021]}.
\bibitem[{Tang et~al.(2019)Tang, Lin \& He}]{Tang2019a}
\bibinfo{author}{Tang, X.}, \bibinfo{author}{Lin, X.}, \& \bibinfo{author}{He,
  F.} (\bibinfo{year}{2019}).
\newblock \bibinfo{title}{{Robust scheduling strategies of electric buses under
  stochastic traffic conditions}}.
\newblock {\it \bibinfo{journal}{Transportation Research Part C: Emerging
  Technologies}\/},  {\it \bibinfo{volume}{105}\/}, \bibinfo{pages}{163--182}.
  \DOIprefix\doi{10.1016/j.trc.2019.05.032}.
\bibitem[{Teng et~al.(2020)Teng, Chen \& Fan}]{Teng2020a}
\bibinfo{author}{Teng, J.}, \bibinfo{author}{Chen, T.}, \&
  \bibinfo{author}{Fan, W.} (\bibinfo{year}{2020}).
\newblock \bibinfo{title}{{Integrated Approach to Vehicle Scheduling and Bus
  Timetabling for an Electric Bus Line}}.
\newblock {\it \bibinfo{journal}{Journal of Transportation Engineering, Part A:
  Systems}\/},  {\it \bibinfo{volume}{146}\/}, \bibinfo{pages}{1--10}.
  \DOIprefix\doi{10.1061/JTEPBS.0000306}.
\bibitem[{TfL(2019{\natexlab{a}})}]{London2019a}
\bibinfo{author}{TfL} (\bibinfo{year}{2019}{\natexlab{a}}).
\newblock \bibinfo{title}{Bus fleet data {\&} audits}.
\newblock \bibinfo{institution}{Public Transport Company of London}. \URLprefix
  \url{https://tfl.gov.uk/corporate/publications-and-reports/bus-fleet-data-and-audits}
  \bibinfo{note}{[last accessed: 02.04.2021]}.
\bibitem[{TfL(2019{\natexlab{b}})}]{London2020a}
\bibinfo{author}{TfL} (\bibinfo{year}{2019}{\natexlab{b}}).
\newblock \bibinfo{title}{{First all-electric double-deck bus route in west
  London to improve air quality}}.
\newblock \bibinfo{institution}{Public Transport Company of London}. \URLprefix
  \url{https://tfl.gov.uk/info-for/media/press-releases/2020/february/-first-all-electric-double-deck-bus-route-in-west-london-to-improve-air-quality}
  \bibinfo{note}{[last accessed: 02.04.2021]}.
\bibitem[{Tschakert(2020)}]{Tschakert2020a}
\bibinfo{author}{Tschakert, W.} (\bibinfo{year}{2020}).
\newblock \bibinfo{title}{Wie gut ist der solaris-stromer?}
\newblock \URLprefix
  \url{https://busfahrt.com/images/stories/testberichte/solarisurbino_e12_0220.pdf}
  \bibinfo{note}{[last accessed: 16.04.2021]}.
\bibitem[{{van Kooten Niekerk} et~al.(2017){van Kooten Niekerk}, van~den Akker
  \& Hoogeveen}]{vanKootenNiekerk2017a}
\bibinfo{author}{{van Kooten Niekerk}, M.~E.}, \bibinfo{author}{van~den Akker,
  J.~M.}, \& \bibinfo{author}{Hoogeveen, J.~A.} (\bibinfo{year}{2017}).
\newblock \bibinfo{title}{{Scheduling electric vehicles}}.
\newblock {\it \bibinfo{journal}{Public Transport}\/},  {\it
  \bibinfo{volume}{9}\/}, \bibinfo{pages}{155--176}.
  \DOIprefix\doi{10.1007/s12469-017-0164-0}.
\bibitem[{Varga et~al.(2020)Varga, Mariasiu, Miclea, Szabo, Sirca \&
  Nicolae}]{Varga2020a}
\bibinfo{author}{Varga, B.~O.}, \bibinfo{author}{Mariasiu, F.},
  \bibinfo{author}{Miclea, C.~D.}, \bibinfo{author}{Szabo, I.},
  \bibinfo{author}{Sirca, A.~A.}, \& \bibinfo{author}{Nicolae, V.}
  (\bibinfo{year}{2020}).
\newblock \bibinfo{title}{{Direct and Indirect Environmental Aspects of an
  Electric Bus Fleet Under Service}}.
\newblock {\it \bibinfo{journal}{Energies}\/},  {\it \bibinfo{volume}{13}\/},
  \bibinfo{pages}{336--348}. \DOIprefix\doi{10.3390/en13020336}.
\bibitem[{VDL(2020)}]{VDL2020a}
\bibinfo{author}{VDL} (\bibinfo{year}{2020}).
\newblock \bibinfo{title}{Aiming for zero. move. together.}
\newblock \URLprefix
  \url{https://www.vdlbuscoach.com/de/produkte/citea-electric/citea-slf-slfa-electric}
  \bibinfo{note}{[last accessed: 02.04.2021]}.
\bibitem[{ViriCiti(2019)}]{ViriCity2019a}
\bibinfo{author}{ViriCiti} (\bibinfo{year}{2019}).
\newblock \bibinfo{title}{4 essential steps to electric bus procurement}.
\newblock \bibinfo{institution}{Provider of a cloud-based monitoring system to
  improve electric vehicle operations}. \URLprefix
  \url{https://viriciti.com/de/blog/4-essential-steps-to-electric-bus-procurement/}
  \bibinfo{note}{[last accessed: 02.04.2021]}.
\bibitem[{Volvo(2020)}]{Volvo2020a}
\bibinfo{author}{Volvo} (\bibinfo{year}{2020}).
\newblock \bibinfo{title}{Volvo 7900 electric articulated}.
\newblock \URLprefix
  \url{https://www.volvobuses.de/de-de/our-offering/buses/volvo-7900-electric/specifications.html}
  \bibinfo{note}{[last accessed: 02.04.2021]}.
\bibitem[{Wang et~al.(2019)Wang, Lixia, Yongzhong, Kang, Liu, Lixia \&
  Yongzhong}]{Wang2019a}
\bibinfo{author}{Wang, J.}, \bibinfo{author}{Lixia, K.},
  \bibinfo{author}{Yongzhong, L.}, \bibinfo{author}{Kang, L.},
  \bibinfo{author}{Liu, Y.}, \bibinfo{author}{Lixia, K.}, \&
  \bibinfo{author}{Yongzhong, L.} (\bibinfo{year}{2019}).
\newblock \bibinfo{title}{{Effects of Working Temperature on Route Planning for
  Electric Bus Fleets Based on Dynamic Programming}}.
\newblock {\it \bibinfo{journal}{Chemical Engineering Transactions}\/},  {\it
  \bibinfo{volume}{76}\/}, \bibinfo{pages}{907--912}.
  \DOIprefix\doi{10.3303/CET1976152}.
\bibitem[{Wehres et~al.(2016)Wehres, Azadeh, Maknoon \& Bierlair}]{Wehres2016a}
\bibinfo{author}{Wehres, U.}, \bibinfo{author}{Azadeh, S.~S.},
  \bibinfo{author}{Maknoon, Y.}, \& \bibinfo{author}{Bierlair, M.}
  (\bibinfo{year}{2016}).
\newblock \bibinfo{title}{{Modeling Uncertainty for a Catenary-free Electrical
  Bus}}.
\newblock In {\it \bibinfo{booktitle}{16th Swiss Transport Research
  Conference}\/}.
\newblock \bibinfo{address}{Ascona, Switzerland}.
\newblock \URLprefix \url{http://www.strc.ch/2016/Wehres_EtAl.pdf}
  \bibinfo{note}{[last accessed: 16.04.2021]}.
\bibitem[{Wei et~al.(2018)Wei, Liu, Ou \& {Kiavash Fayyaz}}]{Wei2018a}
\bibinfo{author}{Wei, R.}, \bibinfo{author}{Liu, X.}, \bibinfo{author}{Ou, Y.},
  \& \bibinfo{author}{{Kiavash Fayyaz}, S.} (\bibinfo{year}{2018}).
\newblock \bibinfo{title}{{Optimizing the spatio-temporal deployment of battery
  electric bus system}}.
\newblock {\it \bibinfo{journal}{Journal of Transport Geography}\/},  {\it
  \bibinfo{volume}{68}\/}, \bibinfo{pages}{160--168}.
  \DOIprefix\doi{10.1016/j.jtrangeo.2018.03.013}.
\bibitem[{Wen et~al.(2016)Wen, Linde, Ropke, Mirchandani \& Larsen}]{Wen2016a}
\bibinfo{author}{Wen, M.}, \bibinfo{author}{Linde, E.}, \bibinfo{author}{Ropke,
  S.}, \bibinfo{author}{Mirchandani, P.}, \& \bibinfo{author}{Larsen, A.}
  (\bibinfo{year}{2016}).
\newblock \bibinfo{title}{{An adaptive large neighborhood search heuristic for
  the Electric Vehicle Scheduling Problem}}.
\newblock {\it \bibinfo{journal}{Computers {\&} Operations Research}\/},  {\it
  \bibinfo{volume}{76}\/}, \bibinfo{pages}{73--83}.
  \DOIprefix\doi{10.1016/j.cor.2016.06.013}.
\bibitem[{{Wiener Linien}(2020{\natexlab{a}})}]{Vienna2020a}
\bibinfo{author}{{Wiener Linien}} (\bibinfo{year}{2020}{\natexlab{a}}).
\newblock \bibinfo{title}{{Die Wiener Öffis in Zahlen}}.
\newblock \bibinfo{institution}{Public Transport Company of Vienna}. \URLprefix
  \url{https://www.wienerlinien.at/web/wiener-linien/die-wiener-Öffis-in-zahlen}
  \bibinfo{note}{[last accessed: 02.04.2021]}.
\bibitem[{{Wiener Linien}(2020{\natexlab{b}})}]{Vienna2020b}
\bibinfo{author}{{Wiener Linien}} (\bibinfo{year}{2020}{\natexlab{b}}).
\newblock \bibinfo{title}{{Neue XL-Gelenkbusse und große Elektrobusse für
  Wien}}.
\newblock \bibinfo{institution}{Public Transport Company of Vienna}. \URLprefix
  \url{https://www.wienerlinien.at/eportal3/ep/contentView.do/pageTypeId/66526/programId/74579/contentTypeId/1001/channelId/-48278/contentId/83284}
  \bibinfo{note}{[last accessed: 02.04.2021]}.
\bibitem[{Xylia et~al.(2017{\natexlab{a}})Xylia, Leduc, Patrizio, Kraxner \&
  Silveira}]{Xylia2017b}
\bibinfo{author}{Xylia, M.}, \bibinfo{author}{Leduc, S.},
  \bibinfo{author}{Patrizio, P.}, \bibinfo{author}{Kraxner, F.}, \&
  \bibinfo{author}{Silveira, S.} (\bibinfo{year}{2017}{\natexlab{a}}).
\newblock \bibinfo{title}{{Locating charging infrastructure for electric buses
  in Stockholm}}.
\newblock {\it \bibinfo{journal}{Transportation Research Part C: Emerging
  Technologies}\/},  {\it \bibinfo{volume}{78}\/}, \bibinfo{pages}{183--200}.
  \DOIprefix\doi{10.1016/j.trc.2017.03.005}.
\bibitem[{Xylia et~al.(2017{\natexlab{b}})Xylia, Leduc, Patrizio, Silveira \&
  Kraxner}]{Xylia2017a}
\bibinfo{author}{Xylia, M.}, \bibinfo{author}{Leduc, S.},
  \bibinfo{author}{Patrizio, P.}, \bibinfo{author}{Silveira, S.}, \&
  \bibinfo{author}{Kraxner, F.} (\bibinfo{year}{2017}{\natexlab{b}}).
\newblock \bibinfo{title}{{Developing a dynamic optimization model for electric
  bus charging infrastructure}}.
\newblock {\it \bibinfo{journal}{Transportation Research Procedia}\/},  {\it
  \bibinfo{volume}{27}\/}, \bibinfo{pages}{776--783}.
  \DOIprefix\doi{10.1016/j.trpro.2017.12.075}.
\bibitem[{Yao et~al.(2020)Yao, Liu, Lu \& Yang}]{Yao2020a}
\bibinfo{author}{Yao, E.}, \bibinfo{author}{Liu, T.}, \bibinfo{author}{Lu, T.},
  \& \bibinfo{author}{Yang, Y.} (\bibinfo{year}{2020}).
\newblock \bibinfo{title}{{Optimization of electric vehicle scheduling with
  multiple vehicle types in public transport}}.
\newblock {\it \bibinfo{journal}{Sustainable Cities and Society}\/},  {\it
  \bibinfo{volume}{52}\/}, \bibinfo{pages}{1--10}.
  \DOIprefix\doi{10.1016/j.scs.2019.101862}.
\bibitem[{Yutong(2020)}]{Yutong2020a}
\bibinfo{author}{Yutong} (\bibinfo{year}{2020}).
\newblock \bibinfo{title}{Yutong u12}.
\newblock \URLprefix \url{https://en.yutong.com/products/U12-europe.shtml}
  \bibinfo{note}{[last accessed: 02.04.2021]}.
\bibitem[{Zhou et~al.(2020)Zhou, Liu, Wei \& Golub}]{Zhou2020b}
\bibinfo{author}{Zhou, Y.}, \bibinfo{author}{Liu, X.~C.}, \bibinfo{author}{Wei,
  R.}, \& \bibinfo{author}{Golub, A.} (\bibinfo{year}{2020}).
\newblock \bibinfo{title}{Bi-objective optimization for battery electric bus
  deployment considering cost and environmental equity}.
\newblock {\it \bibinfo{journal}{{IEEE} Transactions on Intelligent
  Transportation Systems}\/},  (pp. \bibinfo{pages}{2487--2497}).
  \DOIprefix\doi{10.1109/tits.2020.3043687}.

\end{thebibliography}

\end{document}